\documentclass[%
,aip,
jap,%
 ,reprint%
 ,secnumarabic%
,superscriptaddress
,amssymb, amsmath,nobibnotes, floatfix]{revtex4-1}
\usepackage{bm}
\usepackage{braket}
\usepackage{caption}
\usepackage{color}
\usepackage{dsfont}
\usepackage[]{graphicx}
\usepackage{epstopdf}
\usepackage[subrefformat=parens,labelformat=parens]{subfig} 

\newcommand{\bd}{\bm{d}}
\newcommand{\be}{\bm{e}}

\newcommand{\bn}{\bm{n}}

\newcommand{\bu}{\bm{u}}
\newcommand{\bv}{\bm{v}}

\newcommand{\bx}{\bm{x}}

\newcommand{\bD}{\bm{D}}

\newcommand{\bH}{\bm{H}}

\newcommand{\bK}{\bm{K}}
\newcommand{\bM}{\bm{M}}

\newcommand{\bU}{\bm{U}}

\newcommand{\bzero}{\mathbf{0}}

\newcommand{\Figref}[1]{Fig.~\ref{#1}}
\newcommand{\sfigref}[1]{\subref*{#1}}
\newcommand{\sFigref}[1]{Fig.~\sfigref{#1}}

\newcommand{\BDwidth}{.66\columnwidth}

\newcommand{\newtext}[1]{{#1}}

\begin{document}
\title{Helical edge states and topological phase transitions in phononic systems using bi-layered lattices}

\author{Raj Kumar Pal}
\email[Email: ]{raj.pal@aerospace.gatech.edu}
\affiliation{Daniel Guggenheim School of Aerospace Engineering, Georgia Institute of Technology, Atlanta, GA 30332, USA}
\author{Marshall Schaeffer}
\affiliation{George W. Woodruff School of Mechanical Engineering,	Georgia Institute of Technology, Atlanta, GA 30332, USA}
\author{Massimo Ruzzene}
\affiliation{Daniel Guggenheim School of Aerospace Engineering, Georgia Institute of Technology, Atlanta, GA 30332, USA}
\affiliation{George W. Woodruff School of Mechanical Engineering, Georgia Institute of Technology, Atlanta, GA 30332, USA}



\begin{abstract}

We propose a framework to realize helical edge states in phononic systems using two identical lattices with interlayer couplings between them. 
A methodology is presented to systematically transform a quantum mechanical lattice which exhibits edge states to a phononic lattice, thereby 
developing a family of lattices with edge states.  
Parameter spaces with topological phase boundaries in the vicinity of the transformed system are illustrated to demonstrate the robustness to mechanical imperfections. 
A potential realization in terms of fundamental mechanical building blocks is presented for the hexagonal and Lieb lattices. The lattices are composed of passive components and the building blocks are a set of disks and linear springs. 
Furthermore, by varying the spring stiffness, topological phase transitions are observed, 
illustrating the potential for tunability of our lattices. 
\end{abstract}

\keywords{Topological edge states, helical edge states, phononic crystals, topological phase transitions}


\maketitle
\section{Introduction}

Wave control in periodic mechanical systems is a growing area of research with applications in acoustic cloaking, lenses, rectification and 
wave steering. In this regard, topologically protected boundary modes (TPBM) are a recent development in physics 
which opens novel approaches in mechanical systems. 
Starting from the work of Haldane~\cite{haldane1988model}, who predicted the 
possibility of electronic edge states, 
topological edge modes in quantum systems have been the subject of extensive research
due to their immense potential. 
Their existence has since been demonstrated experimentally~\cite{zhang2005experimental} and they are based on 
breaking  time reversal symmetry to achieve one way chiral edge modes between bulk bands. The bulk bands are characterized by a topological invariant called the Chern number.  
Recently, Kane and Mele~\cite{kane2005quantum,kane2005z} discovered topological modes in systems with 
intrinsic spin orbit (ISO) coupling, which exhibit the quantum spin Hall effect.  
This system does not require breaking of time reversal symmetry, is characterized by the $Z2$ topological invariant, and has also 
been demonstrated experimentally~\cite{bernevig2006quantum}. Smith and coworkers~\cite{goldman2012topological,beugeling2012topological} 
studied the effect of next 
nearest neighbor interactions on a number of lattices and demonstrated competition between the helical and chiral edge states 
when both ISO coupling and time reversal symmetry breaking occur.  
These quantum mechanical systems have been extended to other areas of physics as well. 
For example,
Haldane and Raghu~\cite{haldane2008possible} demonstrated boundary modes in electromagnetic systems following Maxwell's relations, which 
have subsequently been widely studied in the context of photonic systems~\cite{lu2014topological,peano2014topological}. 
   
In recent years, numerous acoustic and elastic analogues of quantum mechanical systems have also been developed. 
Prodan and Prodan~\cite{prodan2009topological} demonstrated edge modes in biological structures,  
where time reversal symmetry is broken by Lorentz forces on ions due to 
the presence of weak magnetic fields. Zhang et al.~\cite{zhang2010topological} developed a systematic way to 
analyze eigenvalue problems which break time reversal symmetry by gyroscopic forces. 
Extending this line of work, Bertoldi and coworkers~\cite{wang2015topological} demonstrated chiral edge modes 
in a hexagonal lattice with the Coriolis forces obtained from gyroscopes which rotate at the same frequency as the 
excitation frequency. 
Another example is the recent work of Kariyado and Hatsugai~\cite{kariyado2015manipulation}, 
where Coriolis forces are generated by spinning 
the whole lattice. All these designs involve components rotating at high frequencies, 
or excitation of rotating masses.
There have also been a few mechanical boundary modes, based on modulating the stiffness in time. Examples include the works of 
Khanikaev and coworkers~\cite{khanikaev2015topologically,fleury2015floquet}, Yang et al.~\cite{yang2015topological}, \newtext{Deymier and 
coworkers~\cite{swinteck2015bulk}} and Carusotto et al.~\cite{salerno2015floquet}.  
We remark here that all the above devices require energy input for operation and are based on active components. This need to provide energy is 
an obvious drawback for practical applications. 

On the other hand, a few works have focused on mechanical analogues of the helical edge states that arise from ISO coupling and they typically require passive components.   
S{\"u}sstrunk and Huber~\cite{susstrunk2015observation} demonstrated experimentally these edge states in a square lattice network of double pendulums, 
connected by a complex network of springs and levers.  
Recently, Chan and coworkers~\cite{xiao2015synthetic} numerically demonstrated edge modes in an acoustic system, comprised of stacked hexagonal lattices and 
tubes interconnecting the layers in a chiral arrangement.
\newtext{Deymier and coworkers~\cite{deymier2015torsional} demonstrated non-conventional topology of the band structure in $1D$ systems supporting rotational waves.}
Khanikaev and coworkers~\cite{mousavi2015topologically} demonstrated numerically helical edge modes in perforated 
thin plates by coupling the symmetric and antisymmetric Lamb wave modes.  The edge modes occur at specially connected interfaces between two 
plates and it is not clear how to extend the 
framework to incorporate general building blocks.  
Indeed, there presently exists a paucity of works demonstrating a simple and systematic way to extend these concepts and build new lattices. 

Towards this end, our work aims to develop a framework for designing lattices which support propagation of topologically protected 
helical edge states in mechanical systems. 
We extend the approach presented by S{\"u}sstrunk and Huber~\cite{susstrunk2015observation} for square lattices to general planar lattices in a simple and systematic way, 
thereby developing a family of mechanical lattices. 
This approach yields a mechanical analogue of a quantum mechanical system exhibiting edge states and it provides a specific lattice network topology. 
Starting with this network topology, we define a parameter space with 
topological phase boundaries, illustrating the range of parameter values for robust edge modes. 
Potential realizations of these lattices in terms of fundamental mechanical 
building blocks are also presented. Furthermore, topological phase transitions can be induced in our mechanical lattices 
by varying one of the spring constants, thereby paving the way for designing tunable lattices using the presented framework. 

The paper is organized as follows. 
The next section presents the details of the general framework and required transformations. 
In the section thereafter, two specific examples are introduced.  In the results section band diagrams and numerical simulations of the examples are presented, with the full set of
accompanying mechanics equations and description in the appendix.  The final section summarizes the work and presents potential future research directions.  


\section{Theory: From quantum to classical mechanical systems}\label{TheorySec}
The existence of topologically protected edge states depends on certain characteristics of the lattice band structure, arising from 
the eigenvalue problem.   
In the quantum mechanical case, this eigenvalue problem arises from the Schr\"{o}dinger equation, while in the mechanical case it arises from 
Newton's laws of motion.
We first describe the attributes of a quantum mechanical system with helical edge states, and then derive its mechanical analogue.  
It relies on the existence of two overlapping pairs of Dirac cones. When spin orbital coupling is introduced, the Dirac cones separate and a 
topologically nontrivial band gap forms~\cite{kane2005quantum}.

\begin{figure}
\centering
	\includegraphics[keepaspectratio,width=0.75\columnwidth]{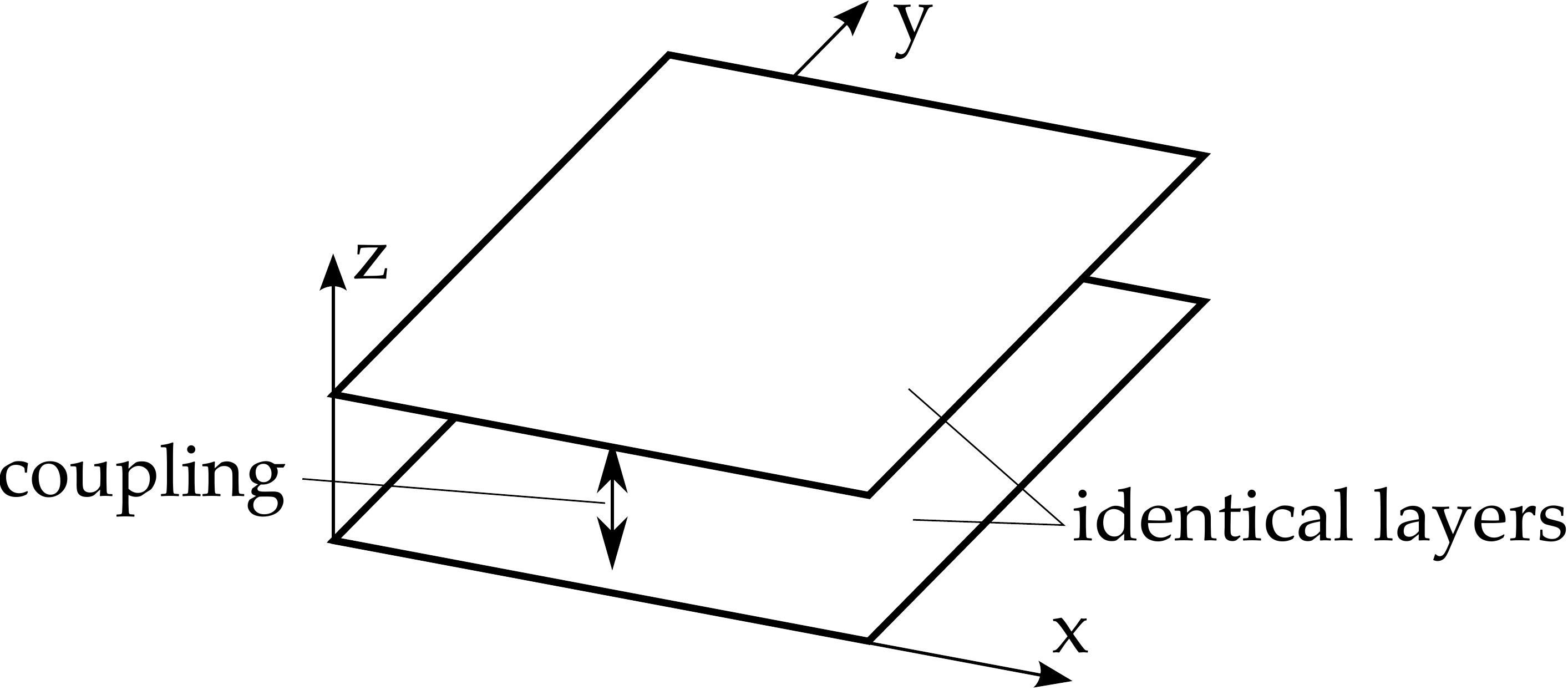}
	\caption{Schematic of bi-layered lattice, made of two identical layers, with interlayer coupling between them.  }
	\label{fig_bilayer_schematic}
\end{figure}

Let $\bH$ be the Hamiltonian of the quantum system for a lattice with up and down spin electrons denoted by 
subscript $\mu$ which takes values in $\{-1,1\}$. 
We consider a lattice with only nearest neighbor interactions and spin orbital coupling, with their interaction strengths denoted by $k_{nn}$
and $\lambda_{iso}$, respectively.   Let $\bd_{pq}$ denote the vector connecting two lattice sites $(p,q)$.
Let $\bn_{pq} = \bd_{pr} \times \bd_{rq} / |\bd_{pr}\times \bd_{rq}|$ denote the unit vector in terms of the bond vectors which connect next nearest neighbor 
lattice sites $p$ and $q$ via the unique intermediate site $r$. 
Let $\sigma_p, p \in\{x,y,z\}$ denote the set of Pauli matrices
and let $\be_z$ denote the unit normal vector in the $z-$direction. 
Following the second neighbor tight binding model~\cite{kane2005quantum} and  assuming that our lattice is located 
in the $xy$ plane, the Hamiltonian for this planar lattice can then be expressed in the form   
\begin{equation}\label{quantumHamilton}
\bH = -k_{nn}\sum_{\Braket{pq} ; \mu} \hat{s}^{\dag}_{p,\mu} \hat{s}_{q,\mu} - 
	i \lambda_{iso} \sum_{\Braket{\Braket{pq}} ; \mu} (\bn_{pq}\cdot \be_z) \sigma_{z,\mu\mu} \hat{s}_{p,\mu}^{\dag} \hat{s}_{q,\mu}  
\end{equation}
Here $\hat{s}_p$ and $\hat{s}^{\dag}_p$ denote the standard creation and annihilation operators, respectively, and $\sigma_{z,\mu\mu}$ equals the value 
of $\mu$.   
The eigenvalue problem arising from the above Hamiltonian has helical edge modes for a range of values of $\lambda_{iso}$ and $k_{nn}$. 

We now describe the procedure to get a mechanical analogue of the above system. In the mechanical system, 
the stiffness matrix comprises of real terms for forces arising from passive components and 
the eigenvalue problem arises from applying a traveling wave assumption to Newton's laws.  
Noting that unitary transformations preserve the eigenvalues and thus the topological properties of the band structures, we apply the following 
unitary transformation~\cite{susstrunk2015observation} to the quantum eigenvalue problem towards obtaining a corresponding Hamiltonian with the 
desired properties in a mechanical system:  
\begin{equation}
\bU = u\otimes \mathds{1}_{N}, \;\;\; u = \dfrac{1}{\sqrt{2}}\begin{pmatrix} 1 & -i \\  1 & i  \end{pmatrix} 
\end{equation} 
where $N$ is the number of sites in the lattice. 
Under this unitary transformation, the mechanical Hamiltonian, is obtained by 
$\bD_H = \bU^{\dag} \bH \bU$. Noting that $u^{\dag}\sigma_z u = i \sigma_y$ and $u^{\dag}u = \mathds{1}_2$, the mechanical Hamiltonian has the following 
expression: 
\begin{equation}\label{dynaEqn}
\bD_H = -k_{nn}\sum_{\Braket{pq}; \mu} \hat{s}^{\dag}_{p,\mu} \hat{s}_{q,\mu}  - 
	\lambda_{iso} \sum_{\Braket{\Braket{pq}}; \mu} \mu (\bn_{pq}\cdot\be_z) s^{\dag}_{p,\mu}  s_{q,-\mu}
\end{equation}
As required, this equivalent Hamiltonian $\bD$ has only real terms and it should arise from the eigenvalue problem of a mechanical system.

Similar to the quantum Hamiltonian, 
the band structure of the dynamical matrix of the corresponding mechanical system requires two overlapping bands, with a bulk band gap between them 
which breaks the Dirac cones and this band gap arises due to the $\lambda_{iso}$ coupling.  
The two overlapping bands are obtained by simply having two identical lattices stacked 
on top of each other, as illustrated in the schematic in Fig.~\ref{fig_bilayer_schematic}. There are masses at the sites.  
The two lattices have nearest neighbor coupling  
springs $k_{nn}$ in their respective planes and springs with strength $\lambda_{iso}$ between the next nearest neighbor sites in the other plane. 
When $\lambda_{iso}=0$, the two lattice layers are uncoupled and the band structure for each layer has two Dirac cones in the 
first Brillouin zone.  The couplings introduced 
by $\lambda_{iso}$  are inter-layer and connect all next nearest neighbors at the lattice site.   
In the presence of couplings between the two lattice layers, the Dirac cones are broken, but the bands still overlap. The band structure is now 
topologically nontrivial and supports edge states similar to the quantum mechanical system.     

Assuming the springs in the lattice to follow a linear force displacement law, Newton's law for displacement $\bu$ at the lattice site $\bx$ takes 
the form $\bM\ddot{\bu} + \bK \bu = \bzero$. Imposing a harmonic solution with frequency $\omega$ 
of the form $\bu(t) = \bv e^{i\omega t}   $ 
results in the eigenvalue problem $\bD \bv = \omega^2 \bv$, where $\bD = \bM^{-1}\bK$ is the dynamical matrix. 
The band structure for an infinite lattice is obtained by imposing the Bloch representation $\bv(\bx) = \bv_0 e^{i\kappa\cdot \bx}$, where $\bv_0$
represents the degrees of freedom in a unit cell of the lattice. 
The eigenvalues of the dynamical matrix $\omega^2$ are positive, in contrast with the eigenvalues of the mechanical Hamiltonian $\bD_H$. 
To have edge modes in the mechanical system with dynamical matrix $\bD$, we require that its band structure  be identical to that of $\bD_H$. 
This condition is achieved by translating the bands of $\bD_H$ upward 
along the frequency axis, so they are all positive. Adding a term $\bD_0 = \eta \mathds{1}_{2N} \;(\eta > 0)$ to $\bD_H$ shifts the bands upward and imposing  
the condition $\bD  = \bD_H + \bD_0$ results in identical band structures. 
Note that adding the term $\bD_0$
keeps the eigenvectors unchanged while adding a constant term $\eta$ to all the eigenvalues, and thus it does not alter the topological properties of the bands. 
Let $n_p$ and $m_p$ denote the number of nearest and next-nearest neighbors at a site $p$.  
Then we have the following expression for $\bD_0$
\begin{equation}
\bD_0 = \sum_{p; \mu} \eta \hat{s}^{\dag}_{p,\mu} \hat{s}_{p,\mu}, \;\;
\eta = \max_{p} \left( k_{nn}m_p + \lambda_{iso} n_p \right). 
\end{equation}
In a mechanical system, the stiffness $k_{pp} = k_{nn}m_p+\lambda_{iso}n_p$ arises at lattice site $p$ when there are springs connecting two 
lattice sites. The dynamical matrix as a result is diagonally dominant and the term $\bD_0$ arises naturally as a result of spring interactions. 
Any additional stiffness required at a lattice site $p$ 
can be achieved by adding a ground spring stiffness equal to the difference between $\eta$ and $k_{pp}$. 
We will demonstrate in the next section that topological phase transitions can arise if the ground spring stiffness deviates from this value.

\section{Mechanical Lattice Description}\label{MechanicalLatticeSection}
\begin{figure}
\centering
\subfloat[]{
	\includegraphics[keepaspectratio,width=0.5\columnwidth]{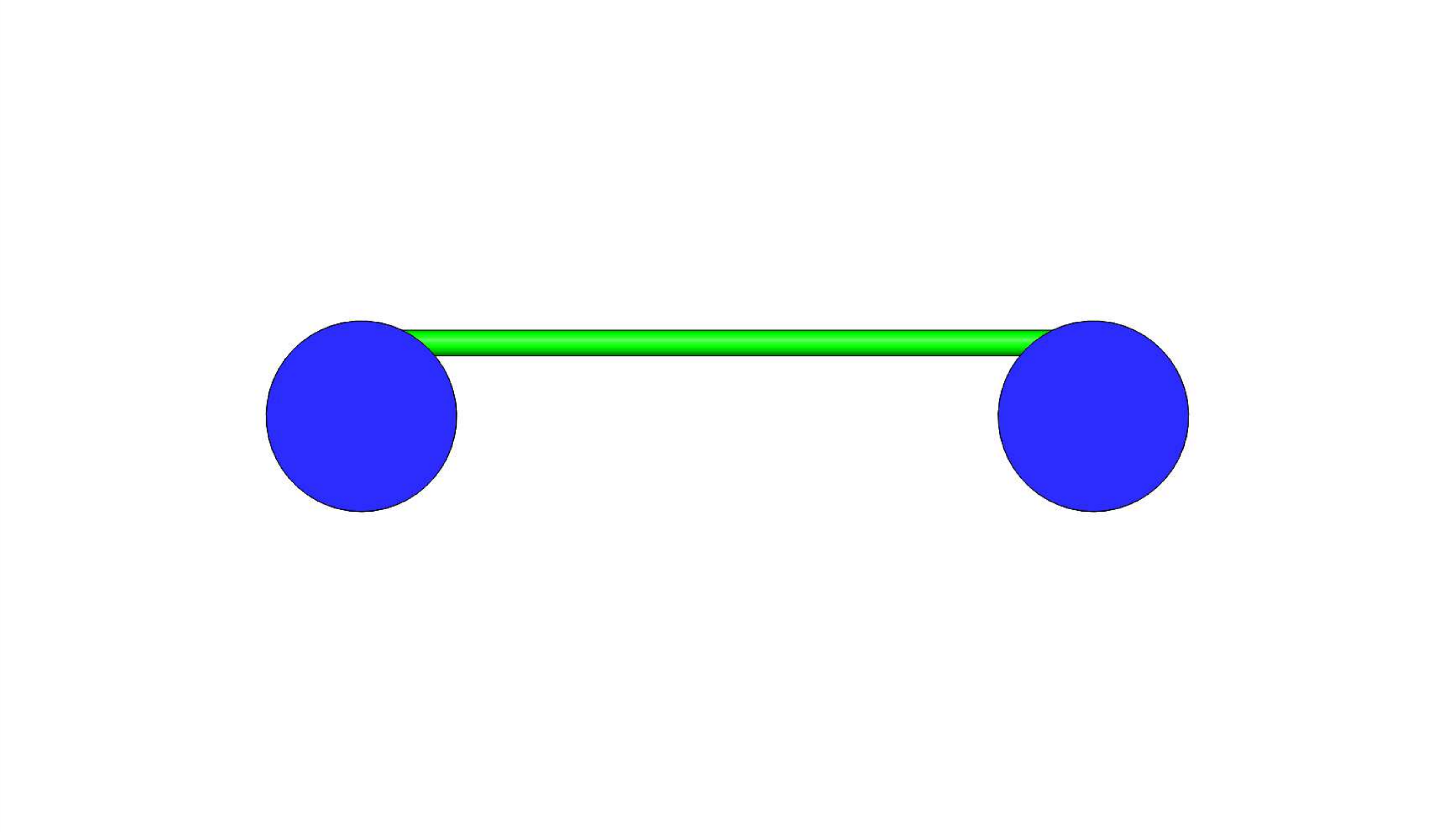}\label{normalSpringSchematic}}
\subfloat[]{
	\includegraphics[keepaspectratio,width=0.5\columnwidth]{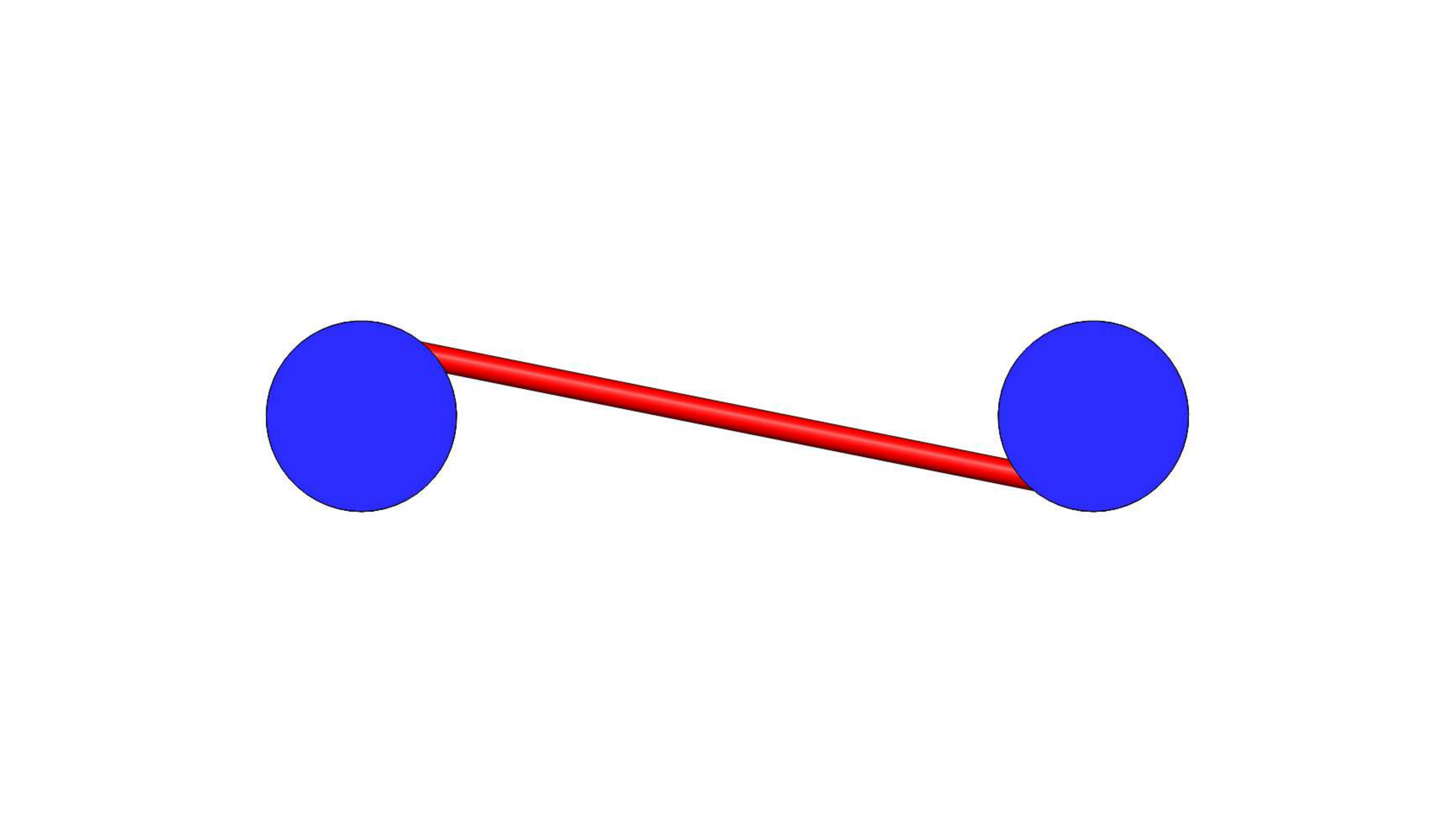}\label{reversedSpringSchematic}}
	\caption{Schematic of (a) normal and (b) reverse springs. A clockwise rotation in one disk induces a clockwise (counter-clockwise) torque 
	on the other from the normal (reverse) spring.}
	\label{springs_schematic}
\end{figure}

\begin{figure}
\centering
\subfloat[]{
	\includegraphics[keepaspectratio,width=0.5\columnwidth]{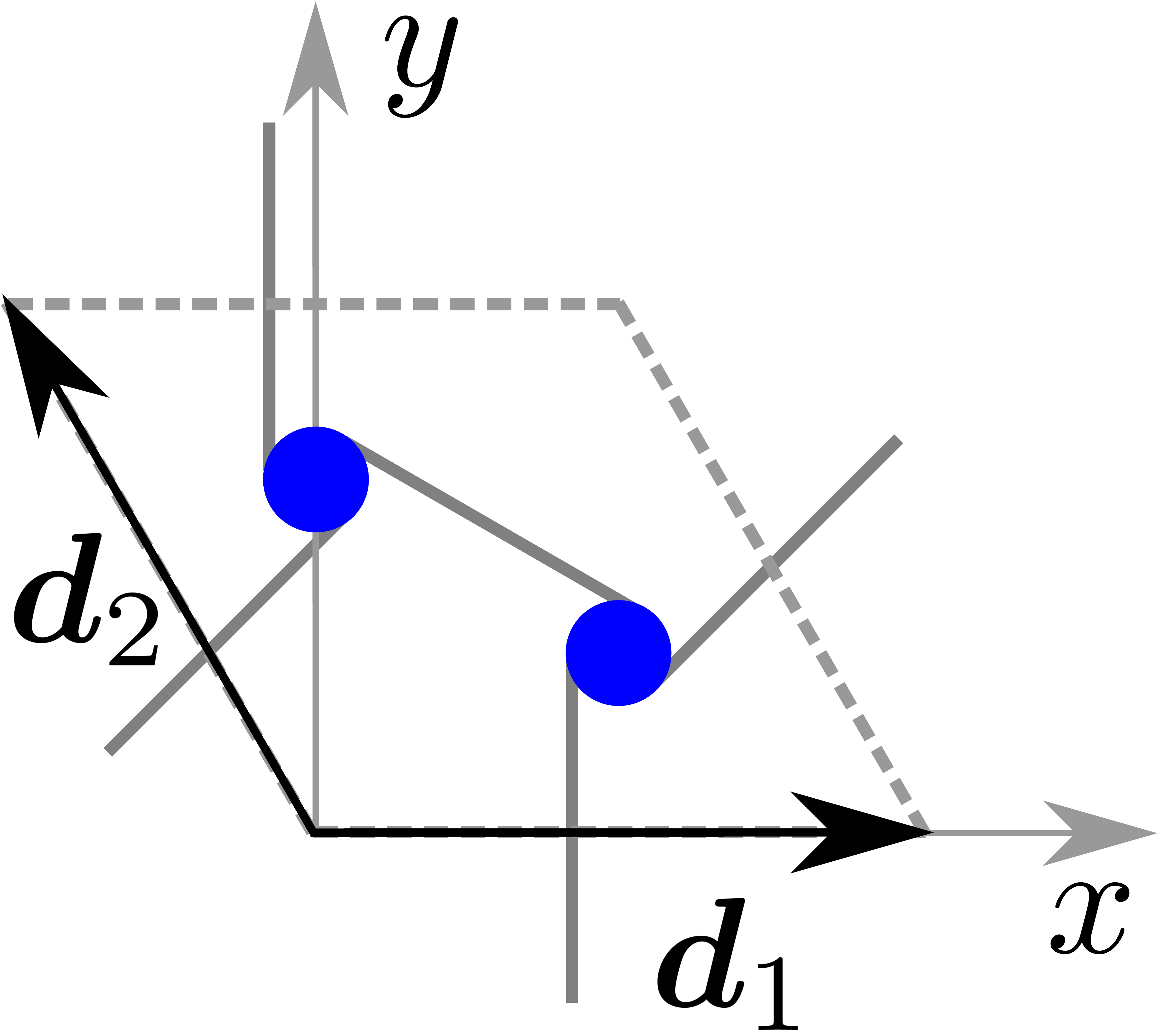}\label{fig_hex_cell}}
\subfloat[]{
	\includegraphics[keepaspectratio,width=0.45\columnwidth]{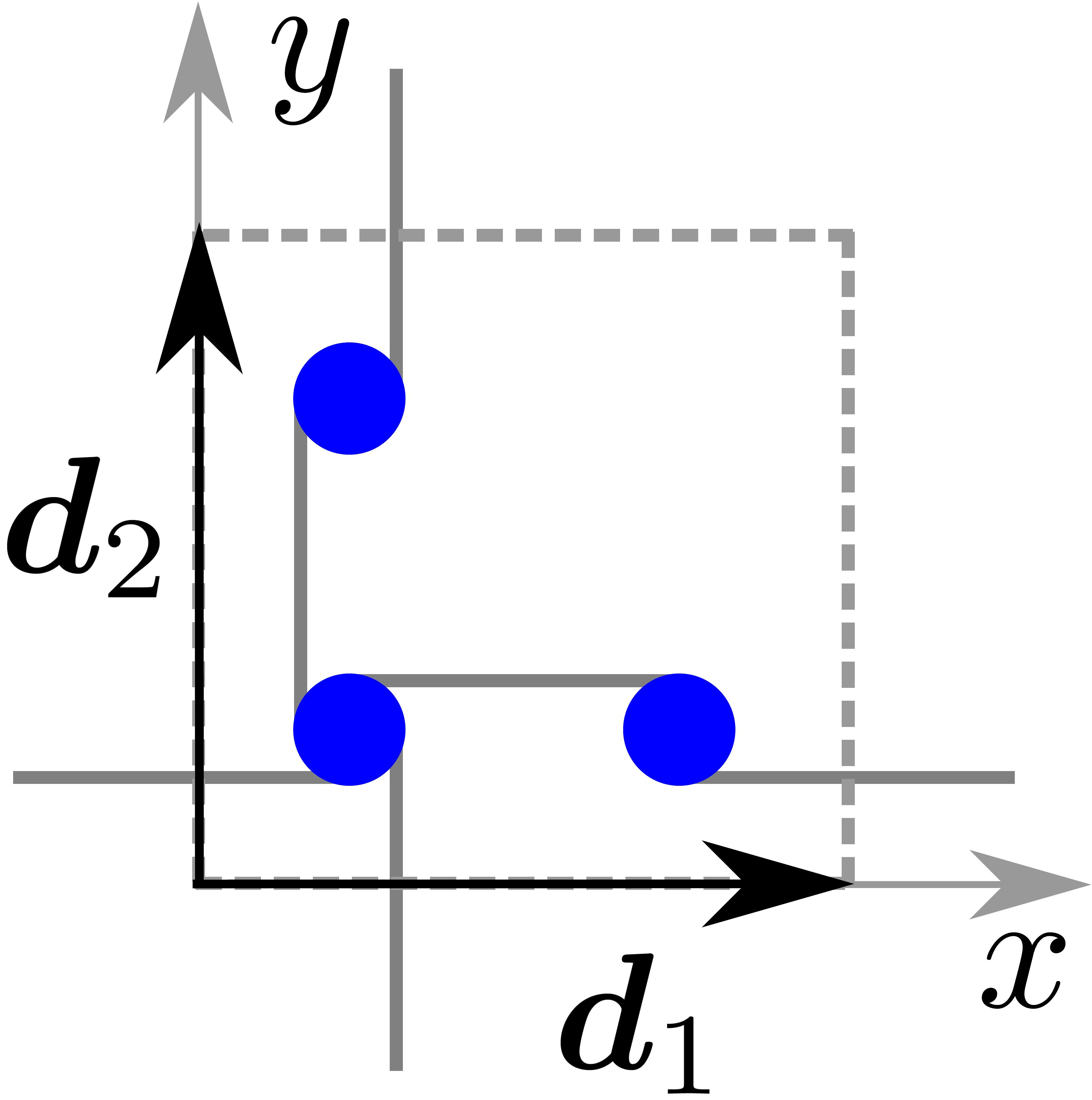}\label{fig_Lieb_cell}}\\
\subfloat[]{
	\includegraphics[keepaspectratio,width=0.5\columnwidth]{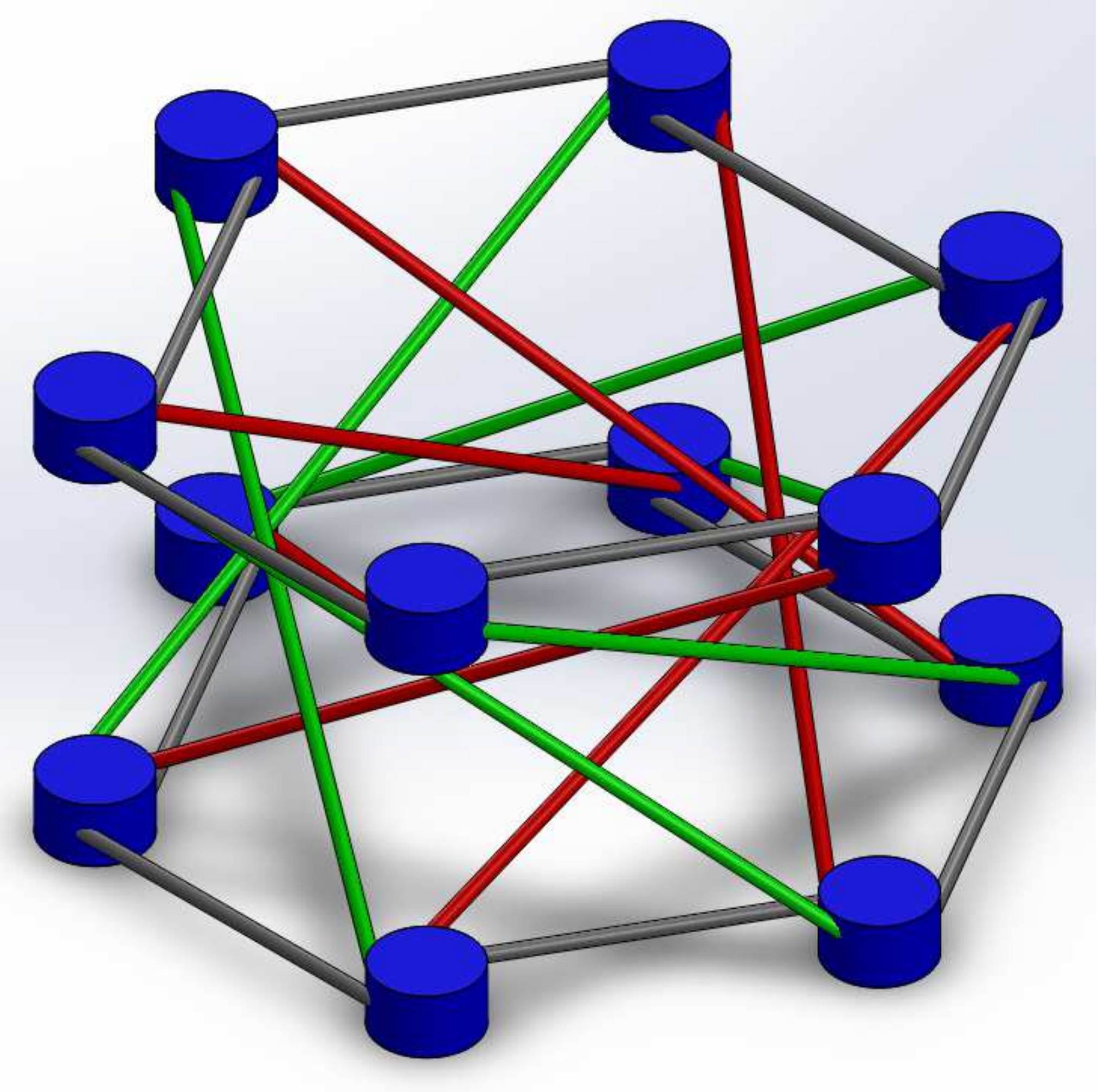}\label{fig_hexTri}}
\subfloat[]{
	\includegraphics[keepaspectratio,width=0.5\columnwidth]{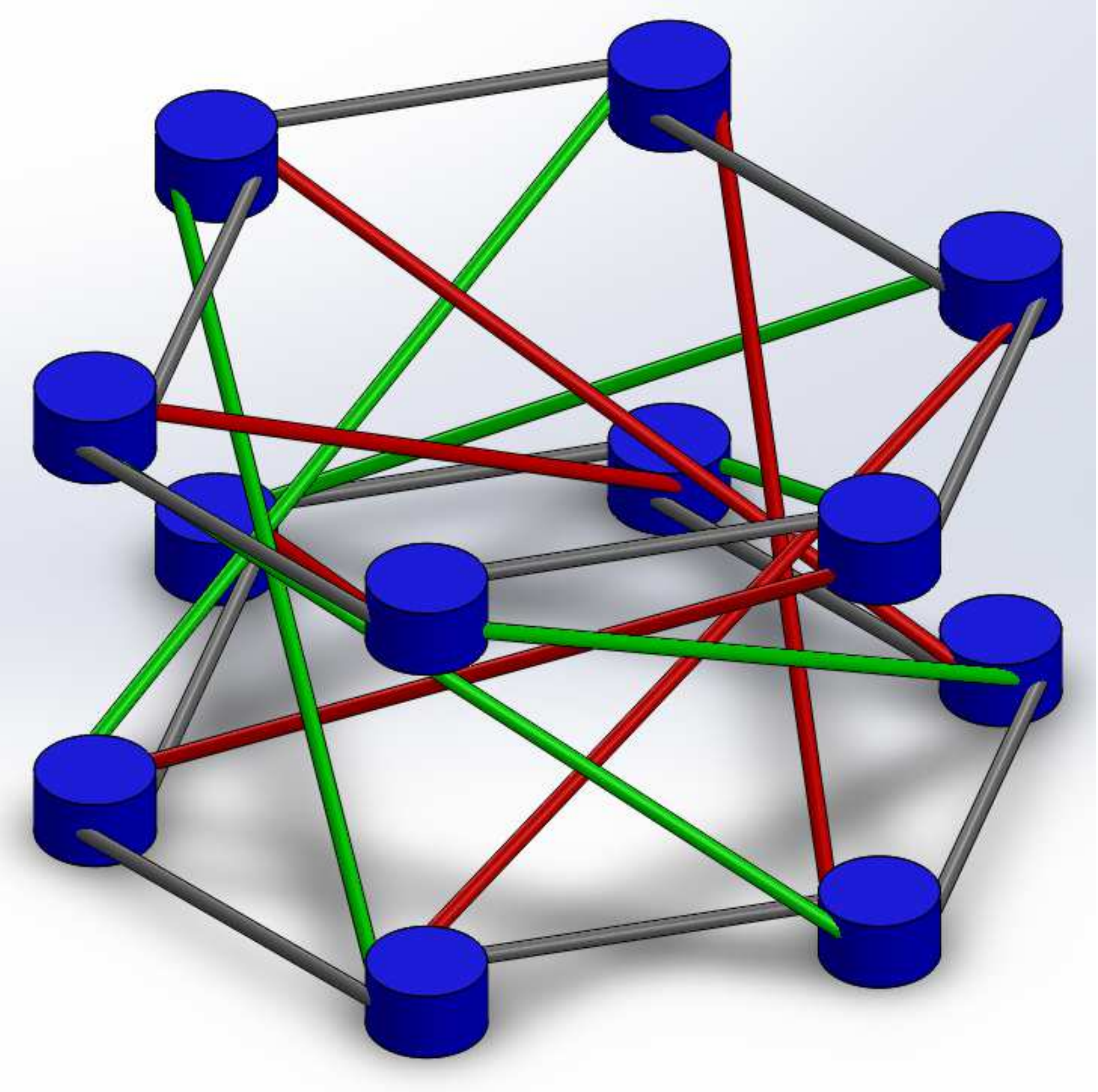}\label{fig_LiebTri}}\\
\subfloat[]{
	\includegraphics[keepaspectratio,width=0.5\columnwidth]{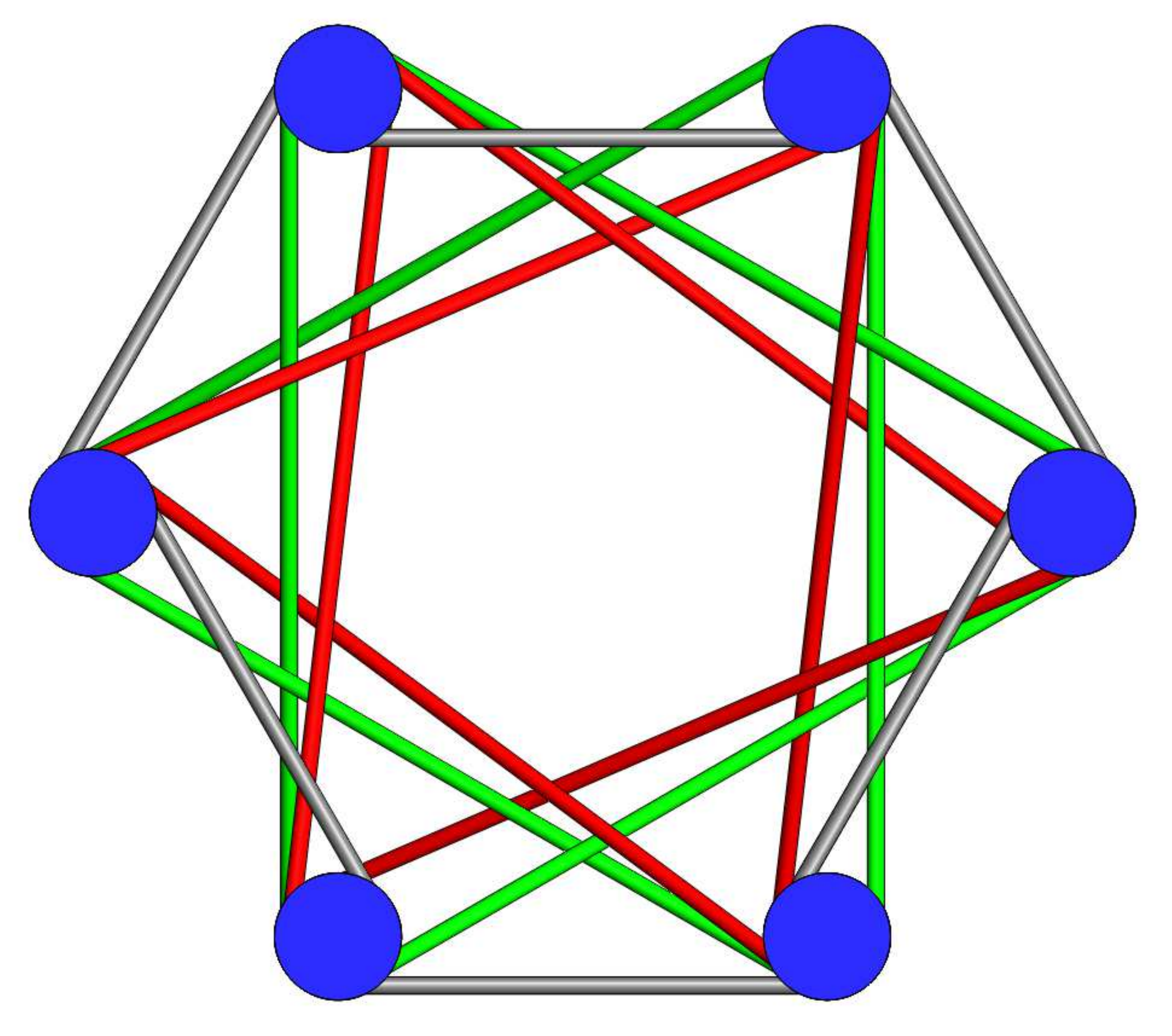}\label{fig_hexTop}}
\subfloat[]{
	\includegraphics[keepaspectratio,width=0.5\columnwidth]{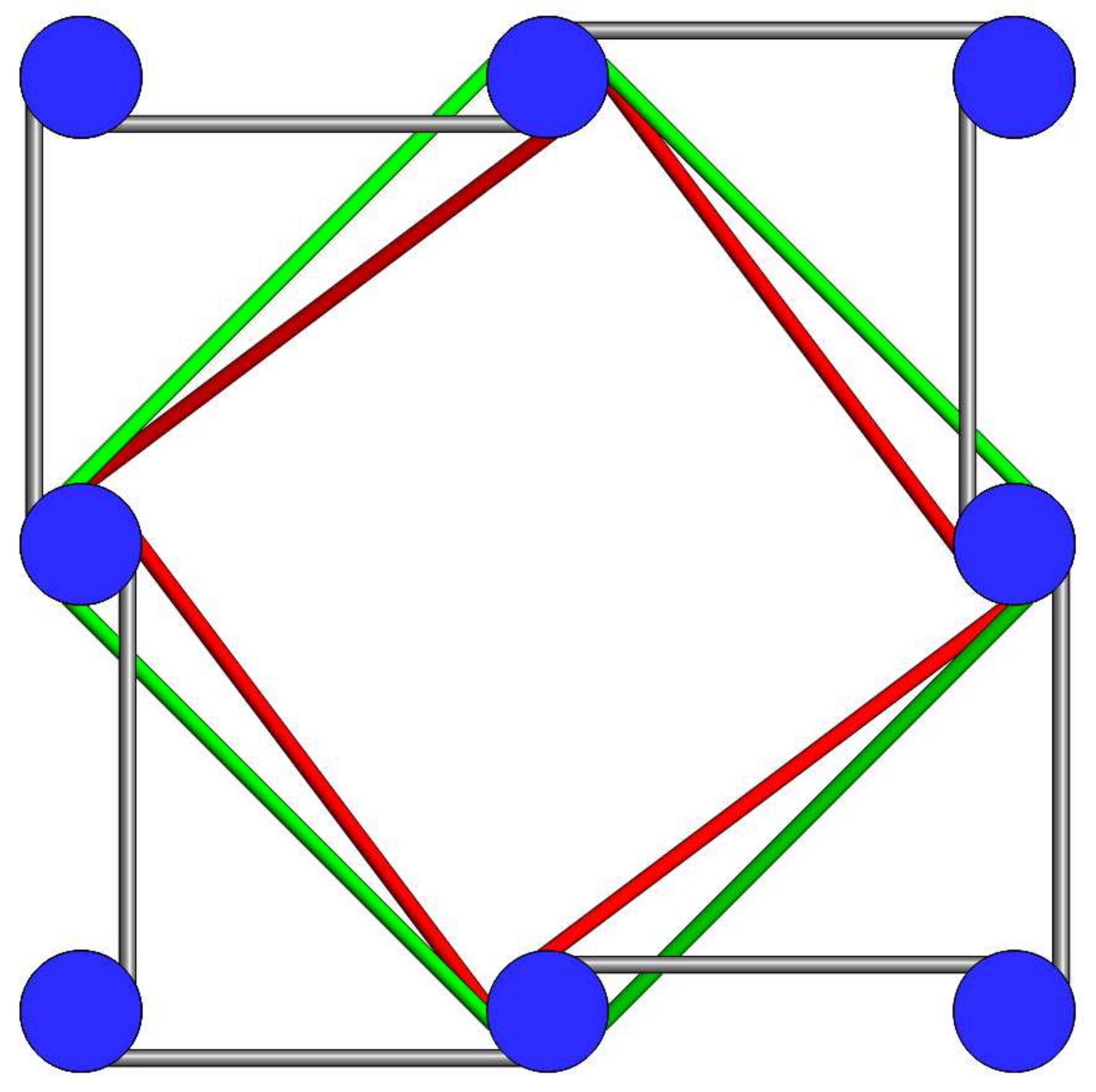}\label{fig_LiebTop}}\\
	\caption{Hexagonal and Lieb lattice respectively (a,b) unit cell defined by the lattice vectors $\bm d_1$ and $\bm d_2$ with nearest-neighbor interactions in gray, (c,d) trimetric view, and (e,f) top view.  (c-f) Bars represent axial springs and the only DOF of each blue cylinder is rotation about its longitudinal axis. Gray bars act within one layer. Red and green bars couple the layers.}
	\label{fig_hex_Lieb_schematic}
\end{figure}


Having presented the structure of a dynamical matrix $\bD$ for realizing edge modes, we now outline the 
procedure to construct a mechanical lattice using fundamental building blocks, which are disks and two types of linear 
springs here called: ``normal'' and ``reversed.''
There is a disk at each lattice site, which can rotate about its center in the $\bm e_z$ direction as its single degree of freedom. It has a rotational inertia $I$ and radius $R$\newtext{, and the mass of the springs relative to the disks are assumed to be negligible}. 
The disks interact with each other by a combination of these normal and reversed springs. Figure~\subref*{normalSpringSchematic} displays a schematic of the normal spring, composed 
of a linear axial spring of stiffness $k_N$ 
connecting the edges of disks $(i,j)$. 
Figure~\subref*{reversedSpringSchematic} displays a schematic of a reversed spring, an axial spring joining opposite sides of a disk. 
For small angular displacements $\theta_i,\theta_j$ the torque exerted on disk $i$ due to the normal and reversed springs are $k_N R^2(\theta_j - \theta_i)$ and 
$k_R R^2(-\theta_j-\theta_i)$, respectively.  \newtext{The aforementioned torque-displacement relations remain a reasonable approximation when the inter-layer springs are nearly parallel to the lattice, which is assumed.  Otherwise, a scale factor can be added.}

We present the explicit construction of two lattice configurations: the hexagonal and the Lieb lattice. Figure~\subref*{fig_hex_cell} displays the schematic of a cell of the hexagonal 
lattice. 
It is composed of two layers of disks, with the disks on the top and bottom layer analogous to the up and down spin electrons at each site in a quantum mechanical system.  
The arrangement of couplings is chosen to achieve the desired form of $\bm D$, with normal in-plane springs (gray color) of stiffness $k_{nn}/R^2$ between nearest neighbors. 
There are two sets of interlayer springs between the next-nearest neighbors, normal springs (green)
and reversed springs (red), both having stiffness $k_{iso}/R^2$. The inter-layer spring between two sites $(p,\mu)$ and $(q,-\mu)$ is of normal or reversed type
depending on the sign of  $\mu (\bn_{pq}\cdot\be_z)$ in Eqn.~\eqref{dynaEqn}. 
Figure~\ref{fig_hex_Lieb_schematic} displays a schematic of the Lieb lattice. Note that the Lieb lattice also requires some normal springs connected to the rigid ground  
due to the mismatch between $\eta$ and $k_{pp}$.  Appendix 1 provides the full set of equations for both the lattices. 

\newtext{In order to practically implement the described mechanical systems a fixture is required to constrain the translation and undesired rotation of the disks.  One can envision a stiff plate or truss above and below the lattice with rods connecting the two through the disks.  As long as the fixture is designed with higher frequency wave modes than the bi-layered lattice it should not interfere with the desired TPBMs.}

\section{Results: Edge Modes and Phase transitions in Bi-layered Mechanical Lattices}\label{ResultsSection}

Bloch wave analysis~\cite{Bloch1929,Brillouin1953,hussein2014dynamics} is conducted to show the existence of TPBMs on lattice boundaries.  
Numerical simulations are used to verify the presence of edge states and demonstrate the excitation of a desired one-way propagating mode. The equations of motion of the lattice 
are solved using a 4th-order Runge-Kutta time-marching algorithm.

\subsection{Mechanical Hexagonal Lattice with TPBMs}\label{sec_Hex}

\begin{figure*}
\centering
\subfloat[]{
	\includegraphics[keepaspectratio,width=\BDwidth]{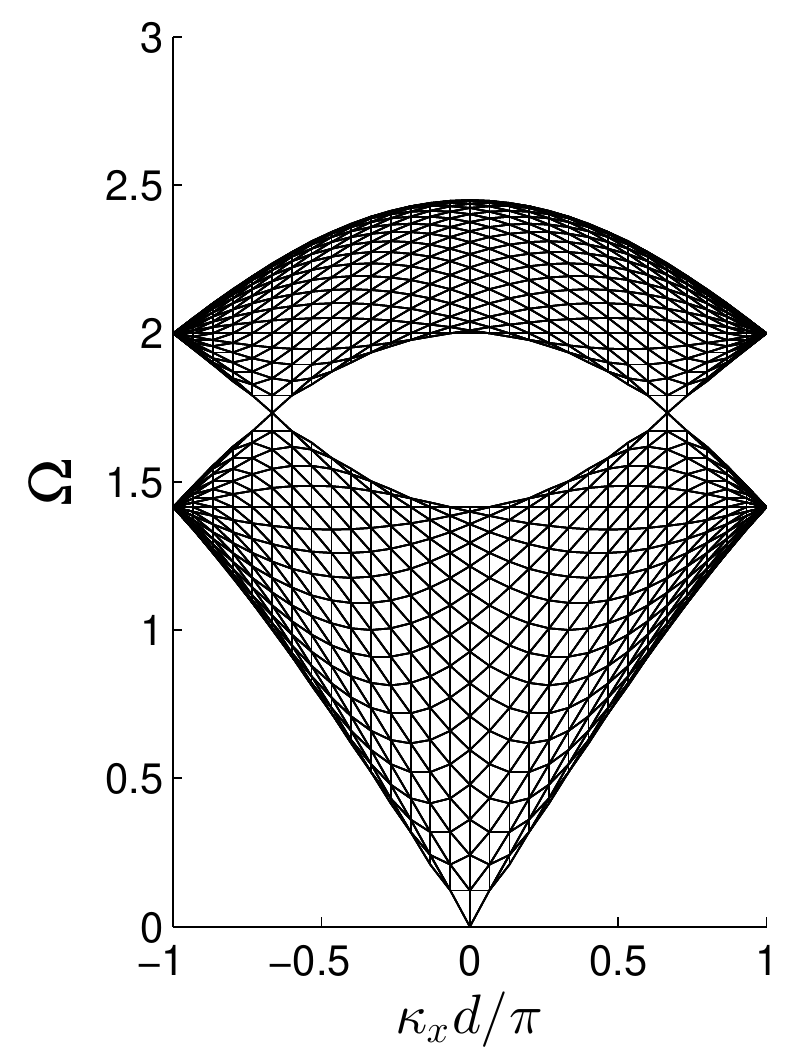}\label{fig_hex_noISO}}
\subfloat[]{
	\includegraphics[keepaspectratio,width=\BDwidth]{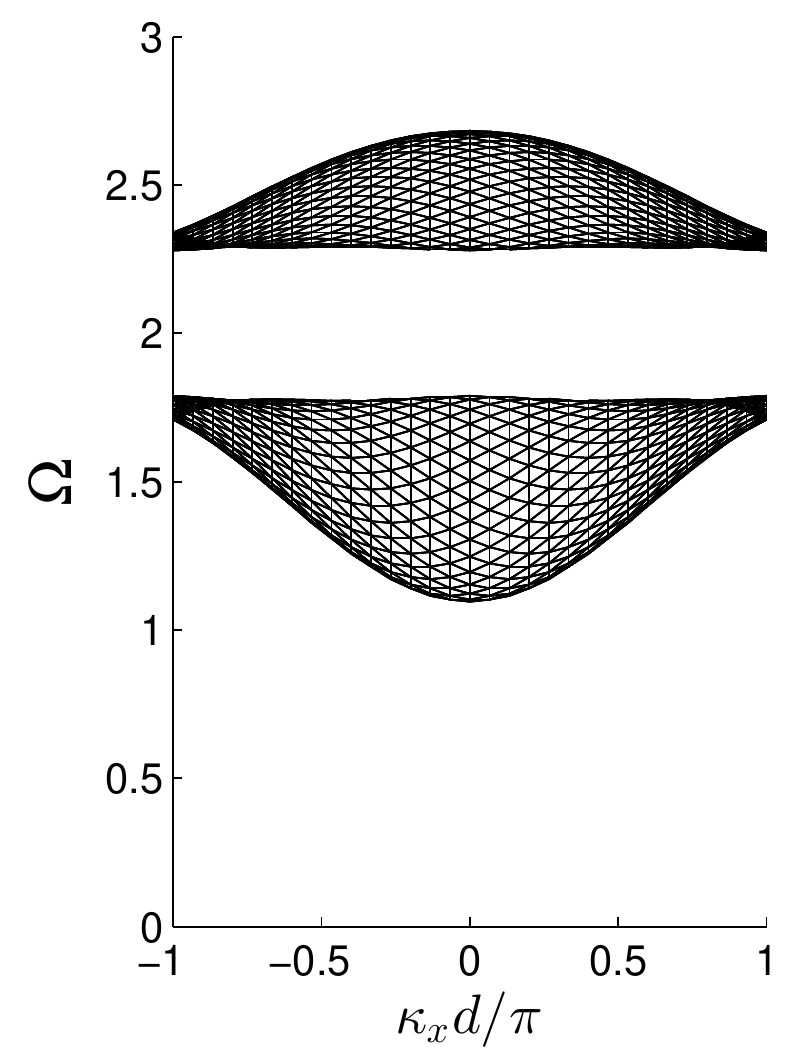}\label{fig_hex_wISO}}
\subfloat[]{
	\includegraphics[keepaspectratio,width=\BDwidth]{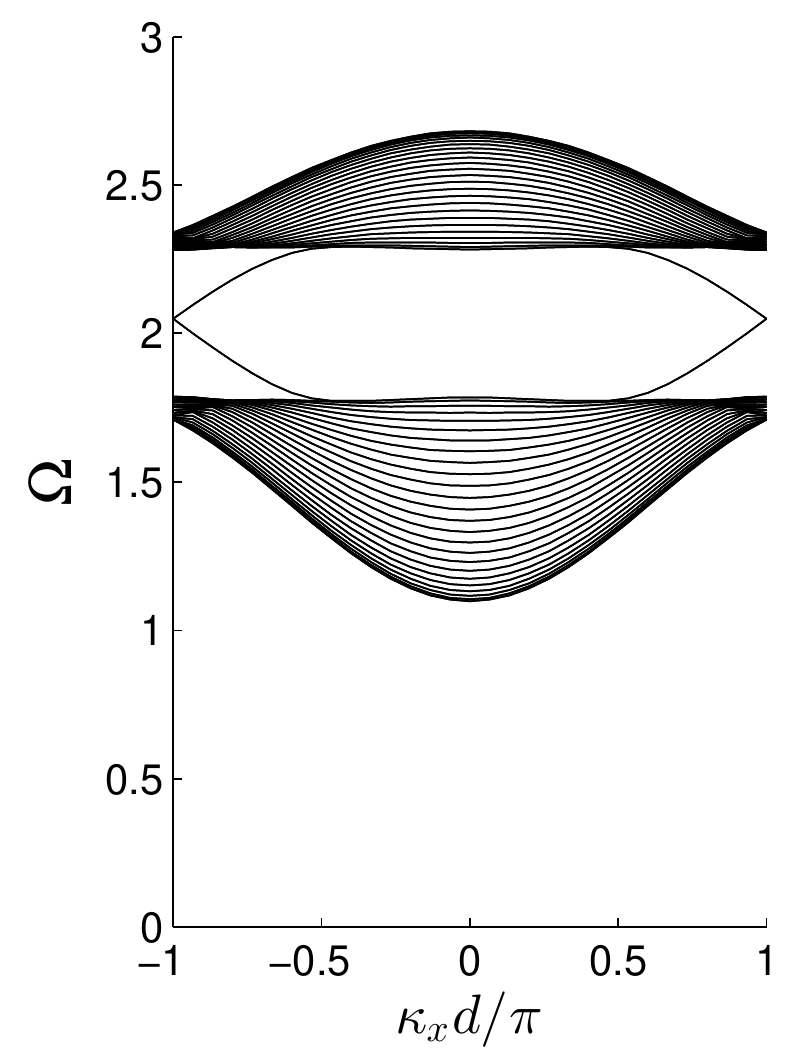}\label{fig_hex_edge}}
	\caption{Bulk band structure with (a) $\lambda_{iso}=0$, (b)$\lambda_{iso}=0.2$, and (c) band diagram of a periodic strip with fixed ends and $\lambda_{iso}=0.2$ coupling.}
	\label{fig_hex_DR}
\end{figure*}

\begin{figure}
\centering
\subfloat[$\tau = 100$]{
	\includegraphics[keepaspectratio,width=0.25\columnwidth]{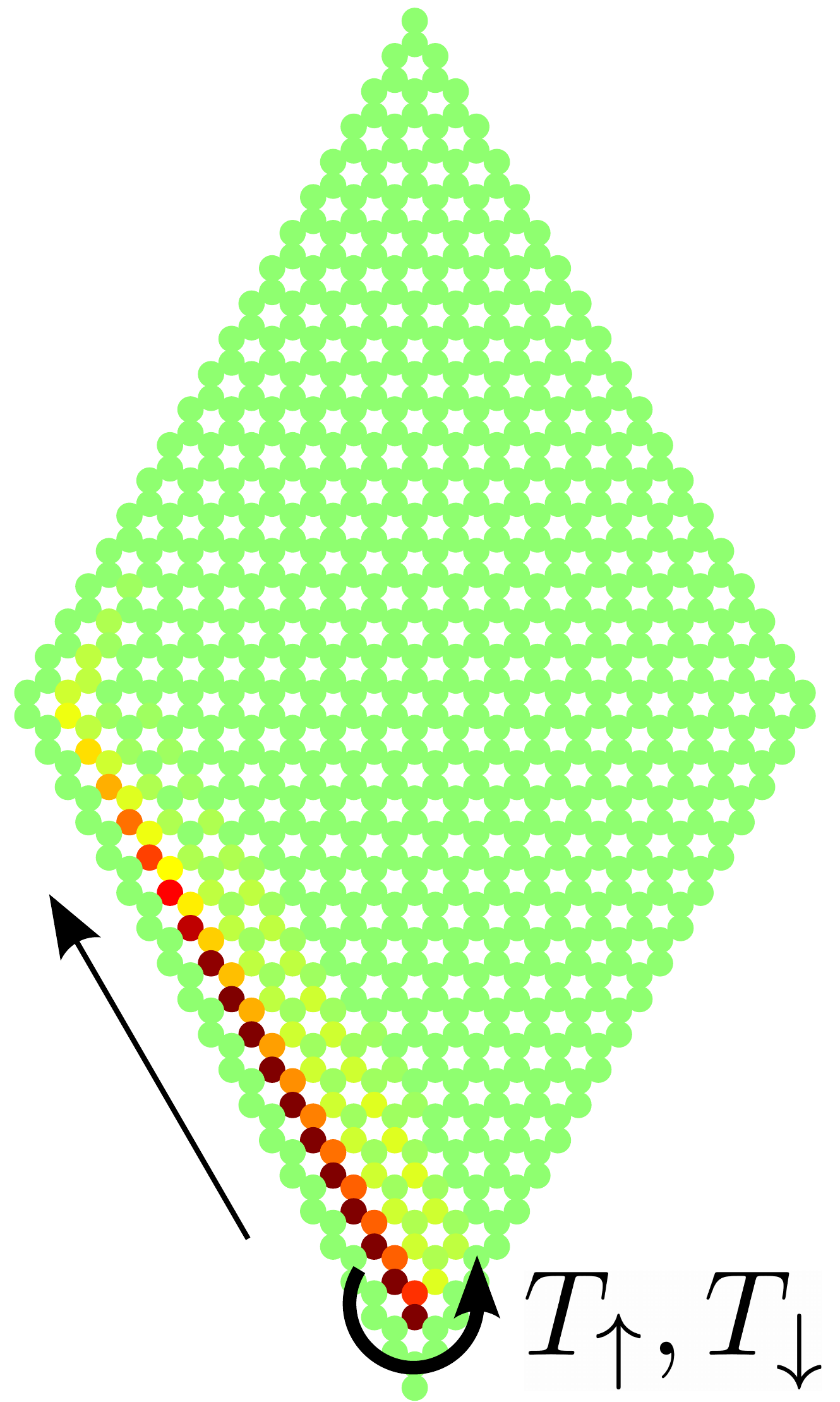}}
\subfloat[$\tau = 200$]{
	\includegraphics[keepaspectratio,width=0.25\columnwidth]{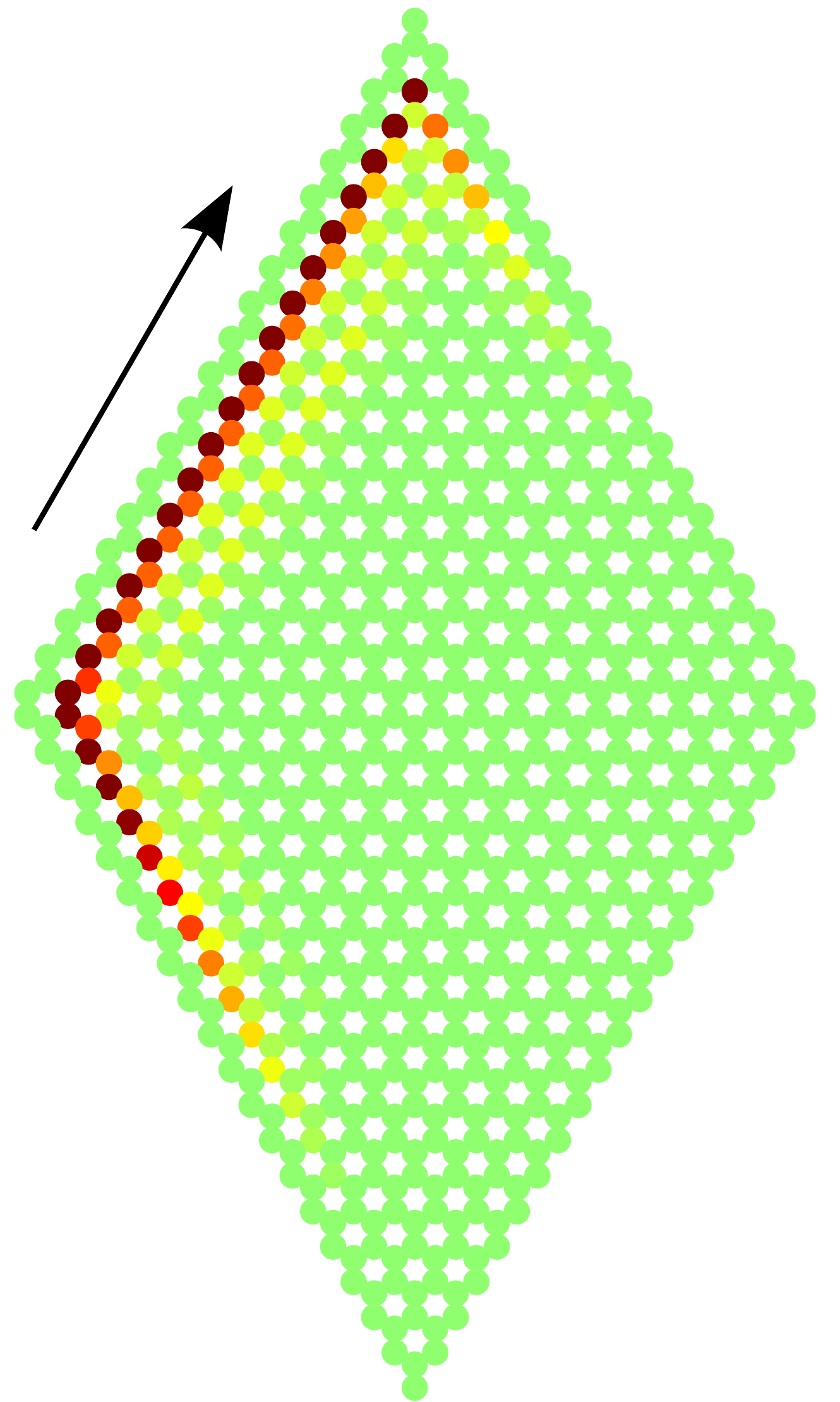}}
\subfloat[$\tau = 300$]{
	\includegraphics[keepaspectratio,width=0.25\columnwidth]{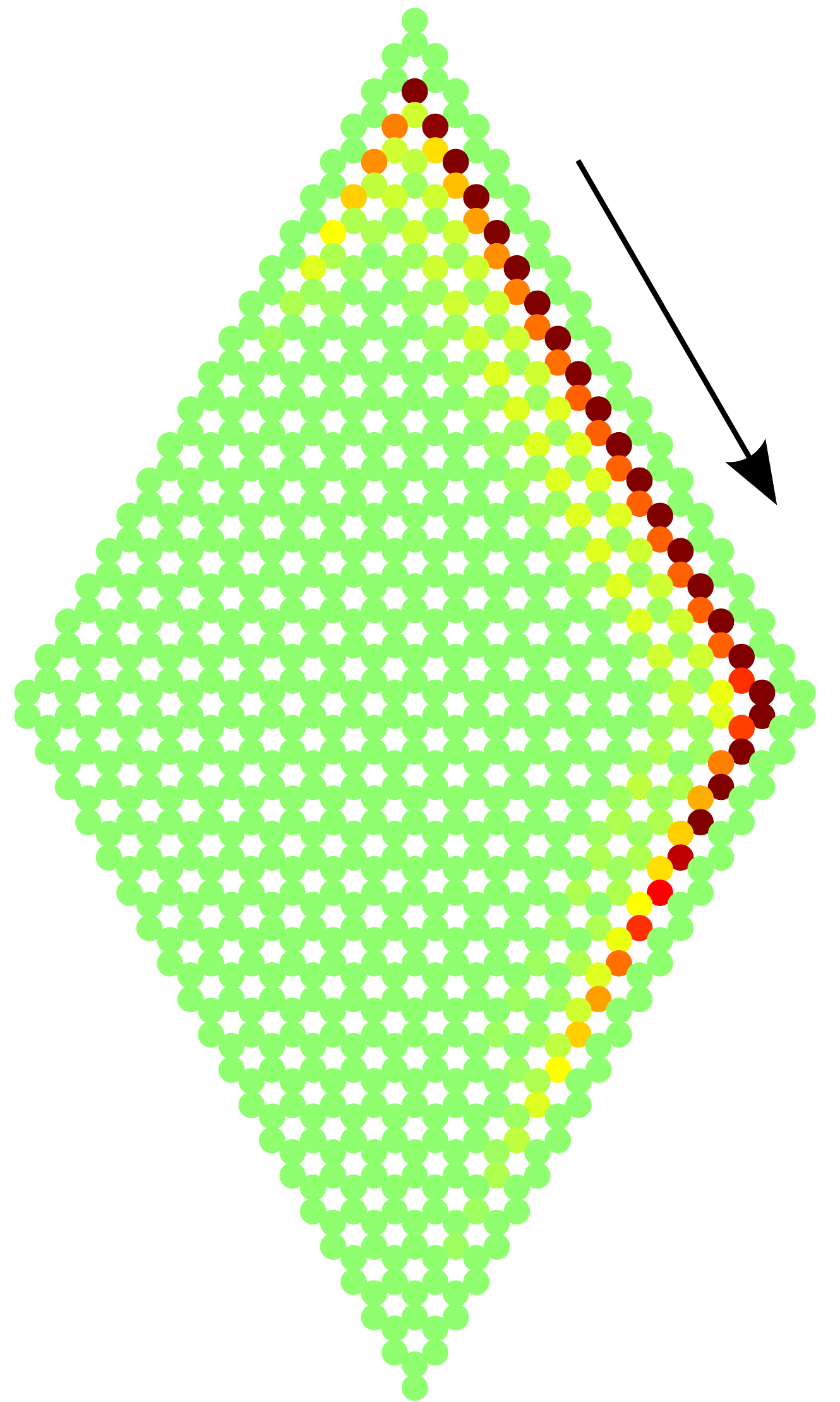}}
\subfloat[$\tau = 400$]{
	\includegraphics[keepaspectratio,width=0.25\columnwidth]{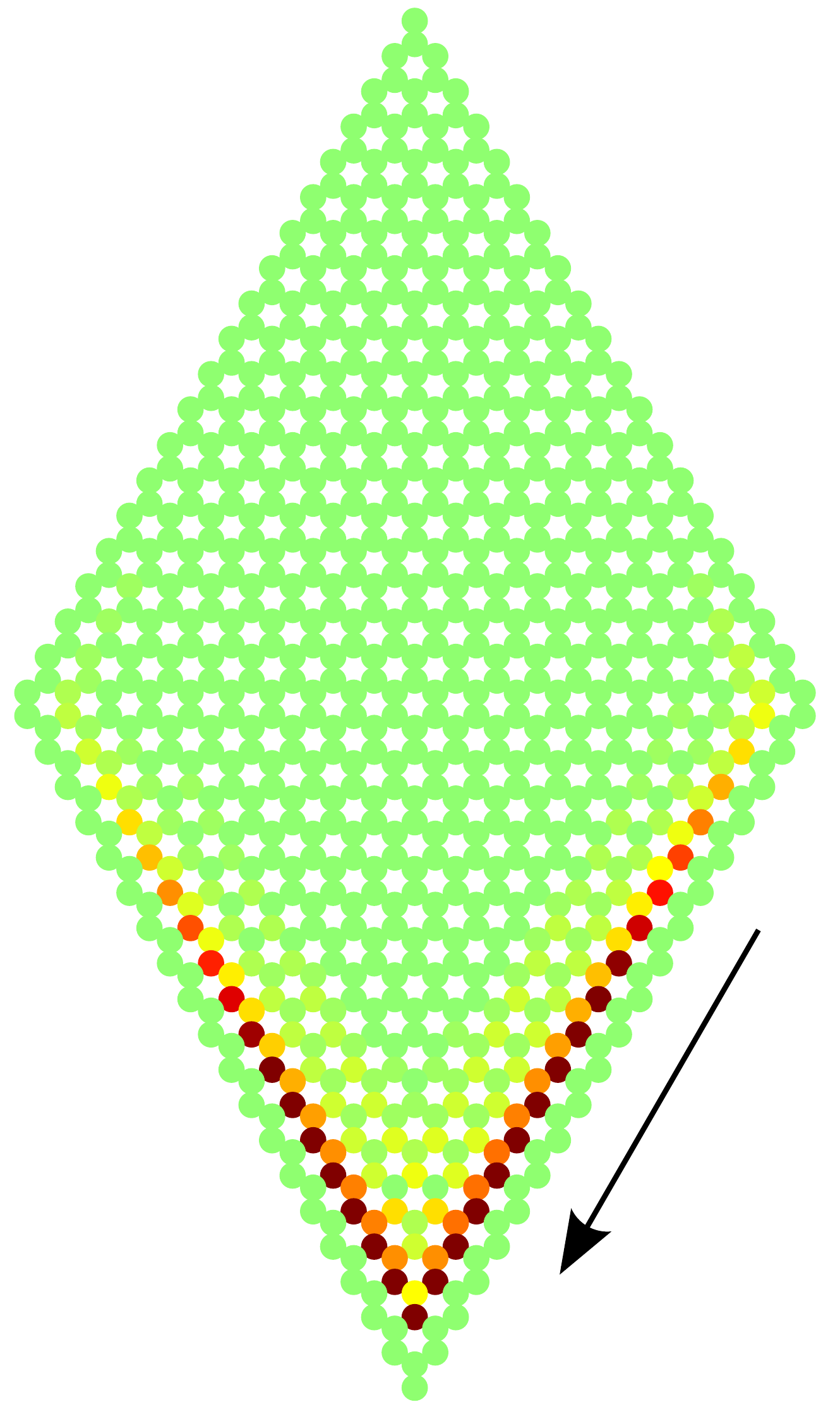}} \\
\subfloat[$\tau = 100$]{
	\includegraphics[keepaspectratio,width=0.25\columnwidth]{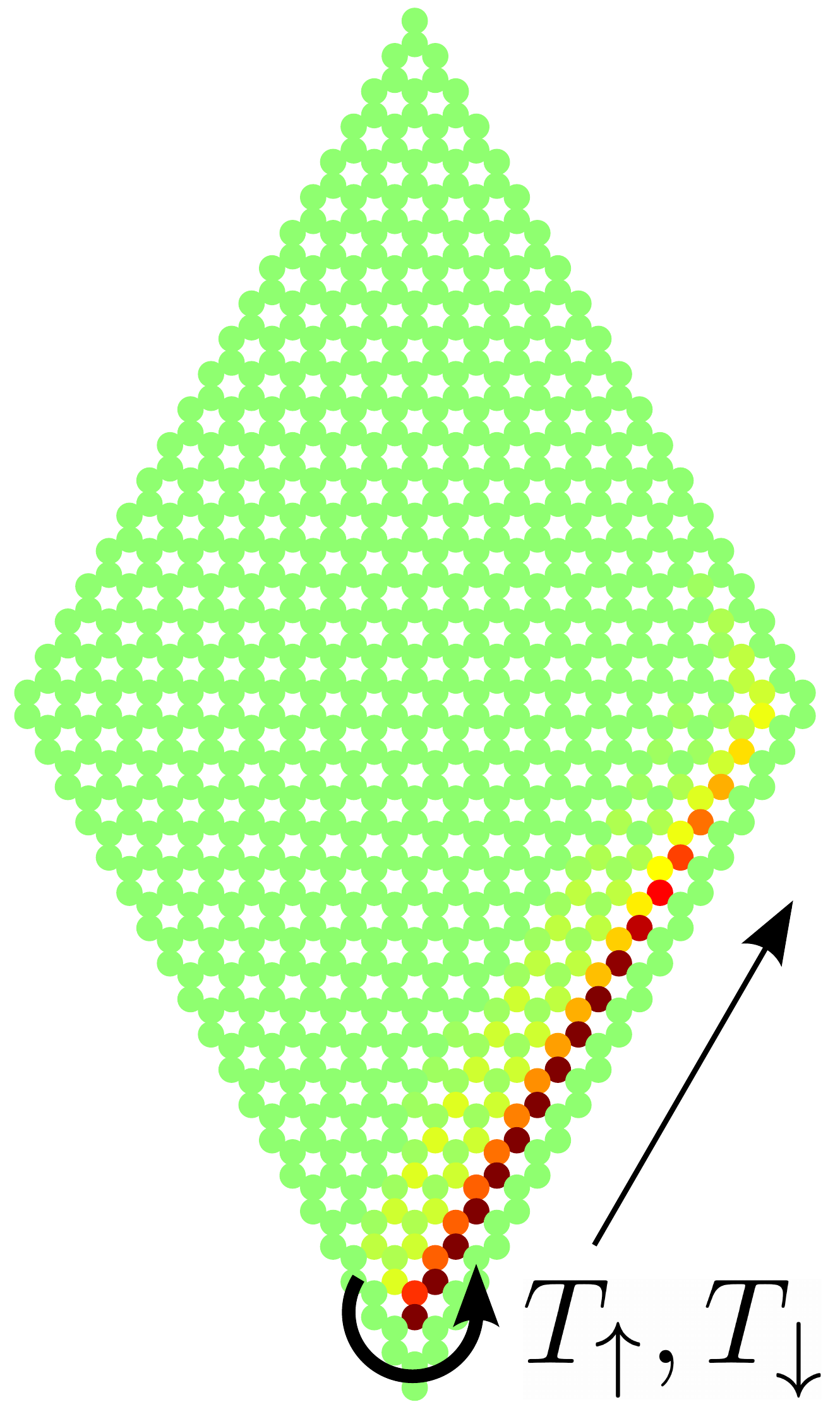}}
\subfloat[$\tau = 200$]{
	\includegraphics[keepaspectratio,width=0.25\columnwidth]{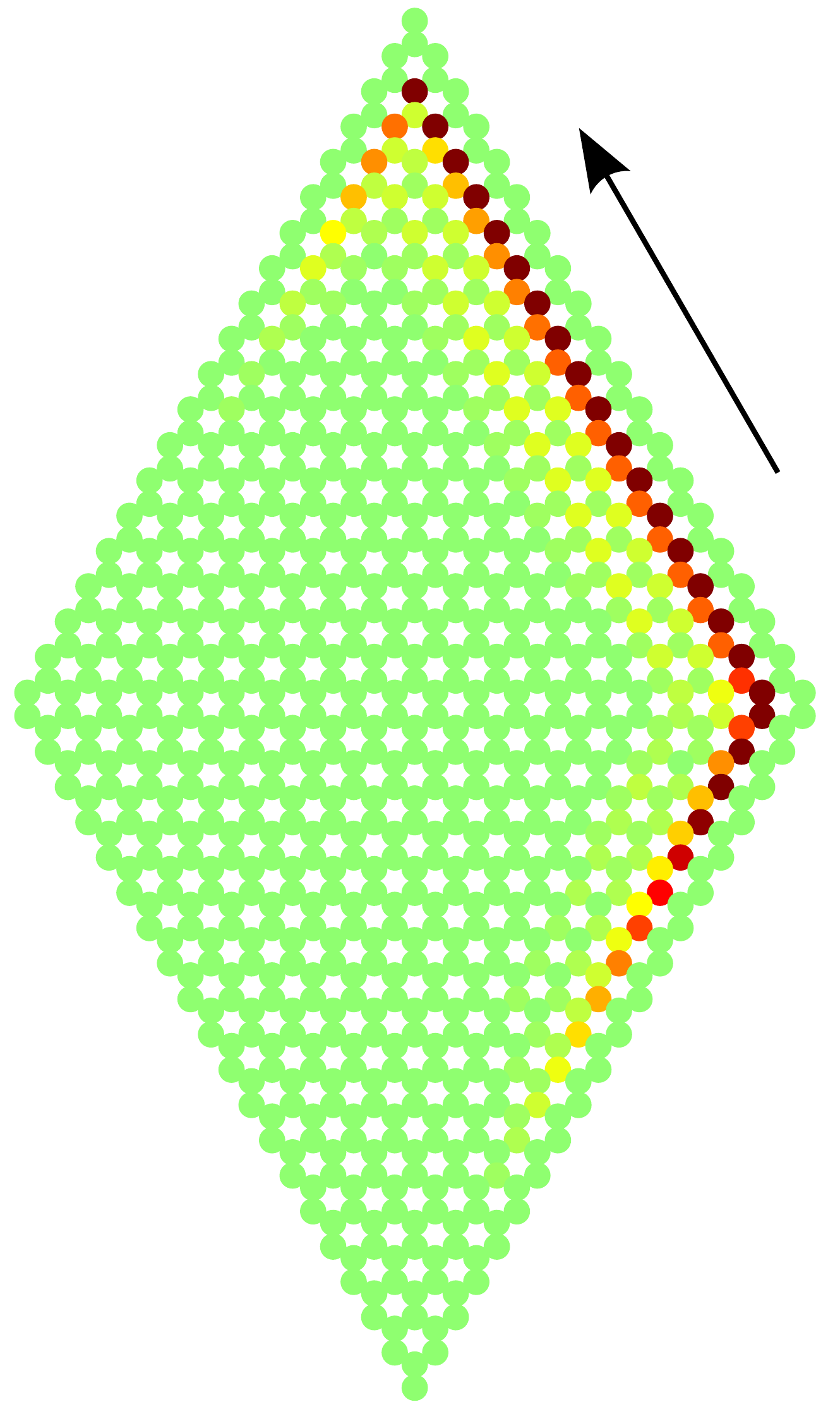}}
\subfloat[$\tau = 300$]{
	\includegraphics[keepaspectratio,width=0.25\columnwidth]{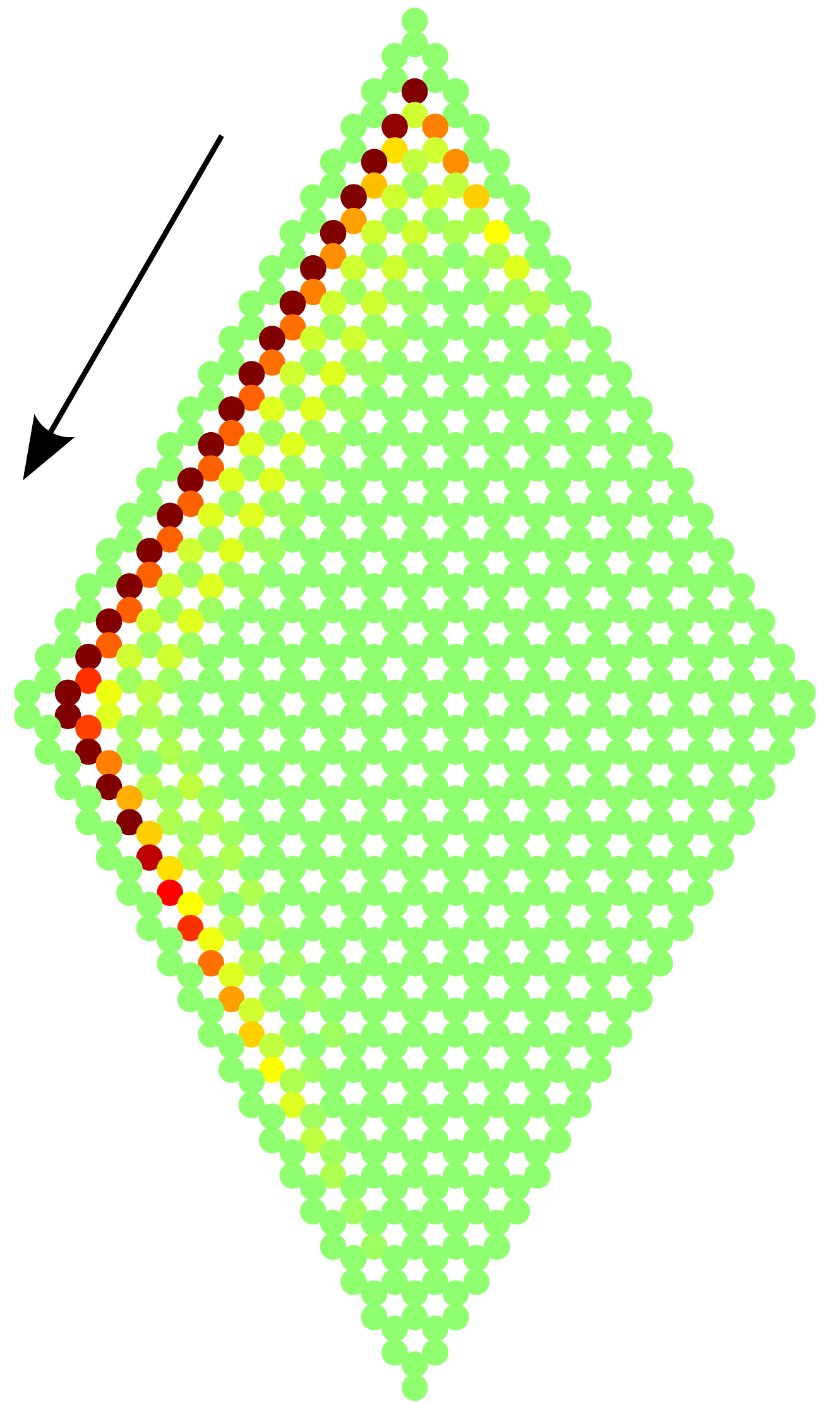}}
\subfloat[$\tau = 400$]{
	\includegraphics[keepaspectratio,width=0.25\columnwidth]{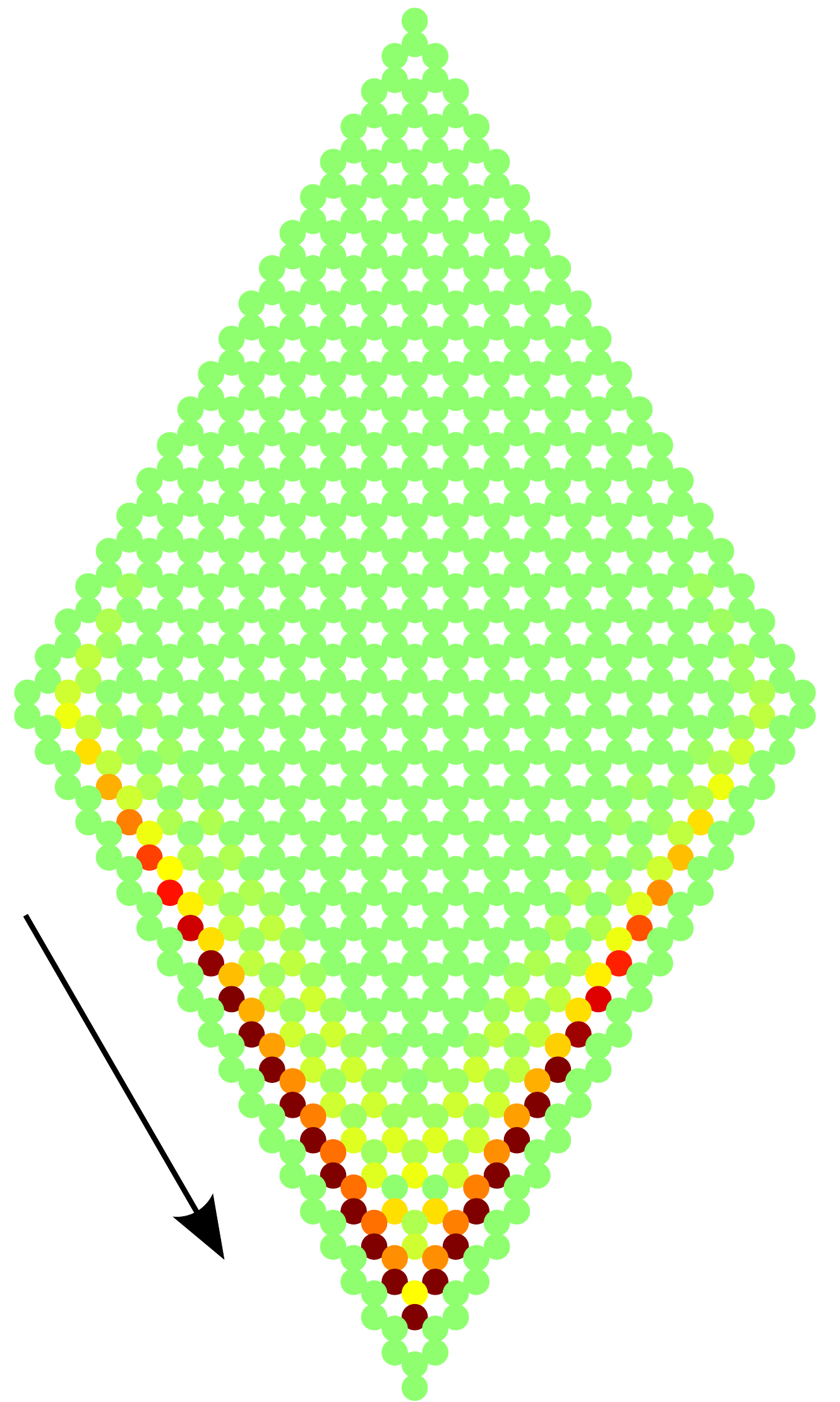}}
	\caption{RMS displacement at each lattice site for different moments in time $\tau$. (a-d) Clockwise propagation.  (e-h) Counter-clockwise propagation.   Red denotes the greatest displacement and green denotes none.}
	\label{fig_hex_NS}
\end{figure}

Figure \ref{fig_hex_DR} displays a projection of the dispersion surface of the hexagonal lattice onto  the $x\Omega$-plane. It illustrates the progression from a structure with no bandgap to one with a topologically nontrivial bulk band gap that exhibits TPBMs.  The wavenumber along $x$, $\kappa_x$, is normalized by the lattice vector magnitude $d = |\bm d_1| = |\bm d_2|$ and $\pi$ for viewing convenience, and $\Omega = \omega/\omega_0$ where $\omega_0^2 = k_{nn}/I$.
Figures \sfigref{fig_hex_noISO} and \sfigref{fig_hex_wISO} are dispersion relations for a 2D, periodic lattice defined by the unit cell in \sFigref{fig_hex_cell}, which approximate the behavior of a lattice far from its boundaries.  Figure \ref{fig_hex_DR}(a) shows the dispersion relation of an uncoupled bilayer system with $k_{nn} = 1$ and $\lambda_{iso} = 0$.  It consists of two identical dispersion surfaces superimposed.  If $\lambda_{iso}$ is instead $0.2$ a topologically nontrivial bandgap opens up in the coupled bi-layered system (\Figref{fig_hex_DR}(b)), so waves at frequencies $\Omega \approx 1.7$ to $2.3$ cannot propagate in the bulk of the lattice, i.e. far from the boundaries.  This is referred to as the ``bulk bandgap''~\cite{lu2014topological}.   Also note that low frequency waves can no longer propagate 
\newtext {and there is a band gap at low frequencies due to the presence of both normal and reverse springs. 
Note that reverse springs prevent zero frequency modes at zero wavenumber. To observe this, consider a chain of disks connected by reverse springs. The equation of motion of 
disk $i$ is $I \ddot{\theta}_i + k_R R^2 (2 \theta_i + \theta_{i+1} + \theta_{i-1}) = 0 $ and the dispersion relation of the system is
$\omega = 2 |\cos(kL/2)|$. They allow low frequency waves only at $k$ close to $\pi$. On the other hand, a chain of disks connected by normal springs have dispersion relation 
$\omega = 2 |\sin(kL/2)|$ and they do not permit low frequency waves close to $\pi$. Hence a combination of these normal and reverse leads to a complete band gap at low frequencies. }

Bloch wave analysis is performed on a strip of the hexagonal lattice to analyze the wave modes localized on the lattice boundaries~\cite{wang2015topological,prodan2009topological} (see Appendix 2 for details).
The disks at each end of the strip are fixed. Figure~\subref*{fig_hex_edge} displays the band diagram, showing $2$ boundary modes spanning the bulk bandgap. Note that there are actually $4$ modes as 
each mode superimposes another mode. On each boundary, there are $2$ modes, propagating in opposite directions and they have different polarizations~\cite{susstrunk2015observation}.  
Thus, an edge excited with a polarization corresponding to a boundary mode will have a wave propagating in only one direction along the boundary.

Numerical simulations are performed to demonstrate the excitation of a TPBM on a finite structure.  The hexagonal lattice unit cell is tessellated into a 20 by 20 cell lattice and the outer layer of cells is fixed.  One site is excited such that the torque is $T_\uparrow = \cos(\Omega\tau)$ and $T_\downarrow = \sin(\Omega\tau)$ on the top and bottom layer respectively, where time $t$ is normalized by defining $\tau = t\omega_0$.  A Hanning window is applied to $T_\uparrow$ and $T_\downarrow$ so only $60$ cycles of forcing are applied, and $\Omega = 2$.  The result is plotted in \Figref{fig_hex_NS}(a-d) for different moments in the simulation.  The coloring of each site corresponds to the root mean square (RMS) of the displacement at that site, 
considering the disks in both the top and bottom layer.  The simulations show that only one wave mode is excited, which propagates in a clockwise direction around the structure, and that this wave mode passes around the two corner types on the lattice boundary without back-scattering.  The displacement magnitude is also seen to decay quickly with increasing distance from the boundary.  There is little dispersion given the relatively constant slope for the edge mode at $\Omega = 2$, so the variation of the amplitude along the boundary is primarily due to the windowing of the excitation.

If the excitation to the lattice is modified so that $T_\downarrow = -\sin(\Omega\tau)$ but $T_\uparrow$ remains the same, only the wave mode which propagates counter-clockwise around the lattice is excited (\Figref{fig_hex_NS}(e-h)).  Thus, the direction of energy propagation can be selected via the polarization of the input to the system.

\subsection{Mechanical Lieb Lattice with TPBMs}
\label{sec_Lieb}

\begin{figure*}
\centering
\subfloat[]{
	\includegraphics[keepaspectratio,width=\BDwidth]{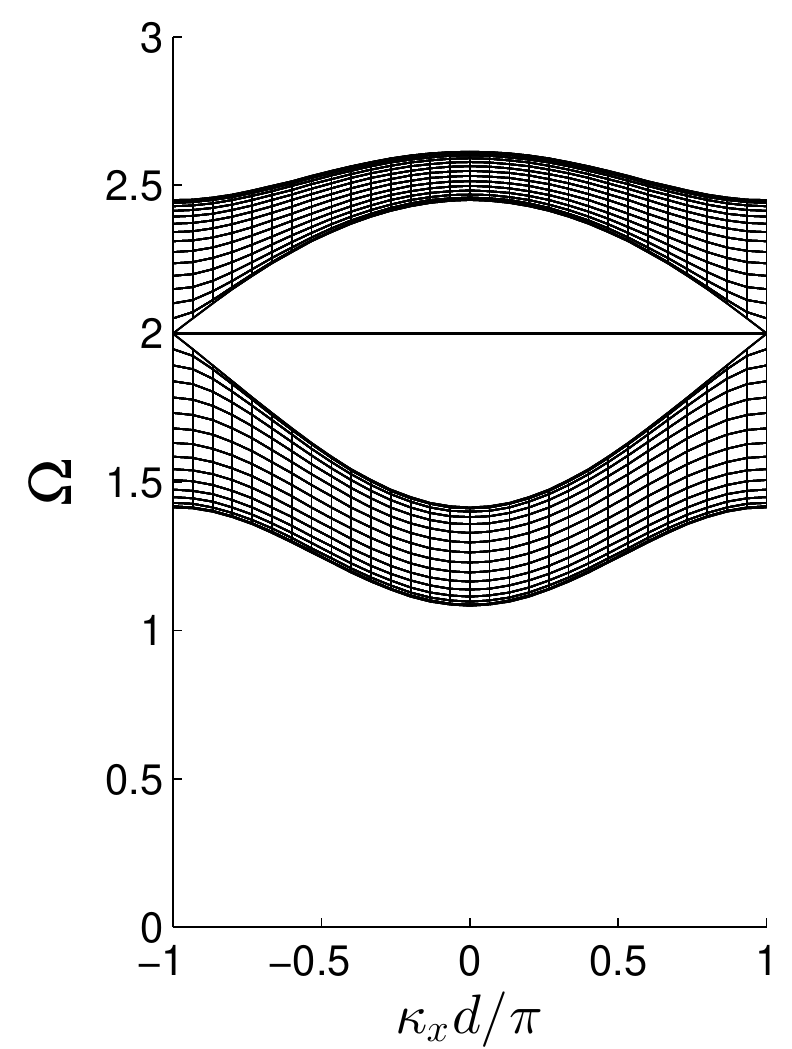}\label{fig_Lieb_noISO}}
\subfloat[]{
	\includegraphics[keepaspectratio,width=\BDwidth]{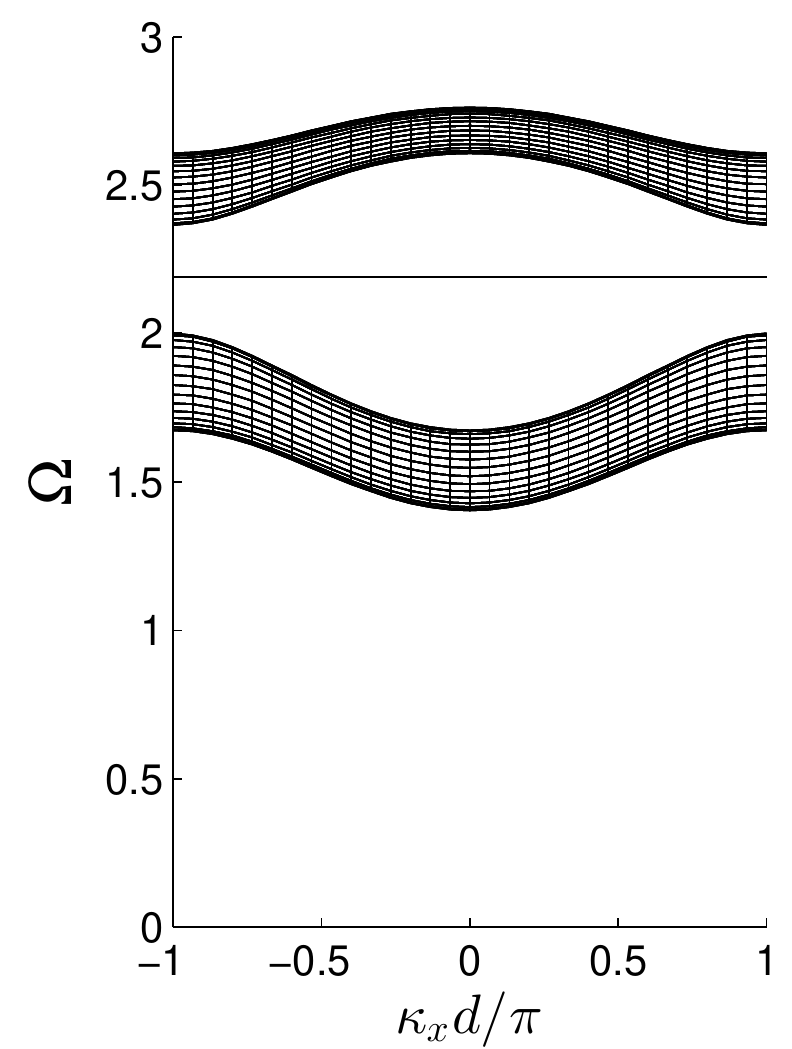}\label{fig_Lieb_wISO}}
\subfloat[]{
	\includegraphics[keepaspectratio,width=\BDwidth]{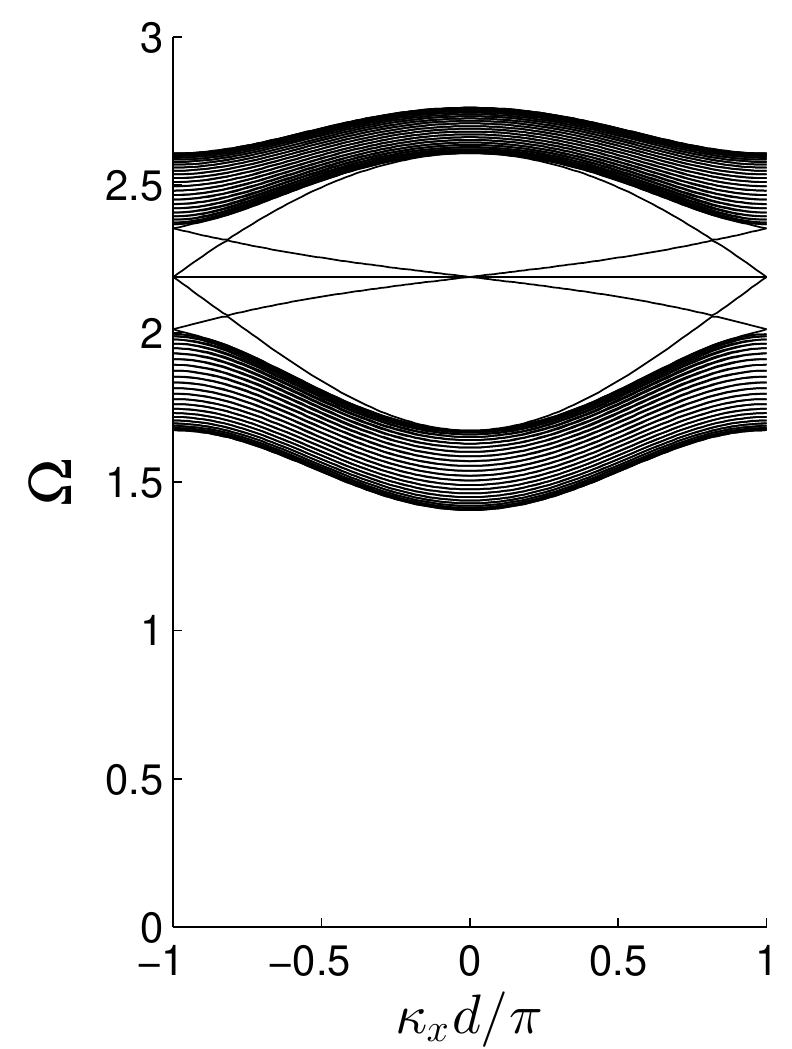}\label{fig_Lieb_edge}}
	\caption{Bulk band structure with (a) $\lambda_{iso}=0$, (b)$\lambda_{iso}=0.2$, and (c) band diagram of a periodic strip with fixed ends and $\lambda_{iso}=0.2$ coupling.}
	\label{fig_Lieb_DR}
\end{figure*}

\begin{figure}
\centering
\subfloat[$\tau = 300$]{
	\includegraphics[keepaspectratio,width=0.33\columnwidth]{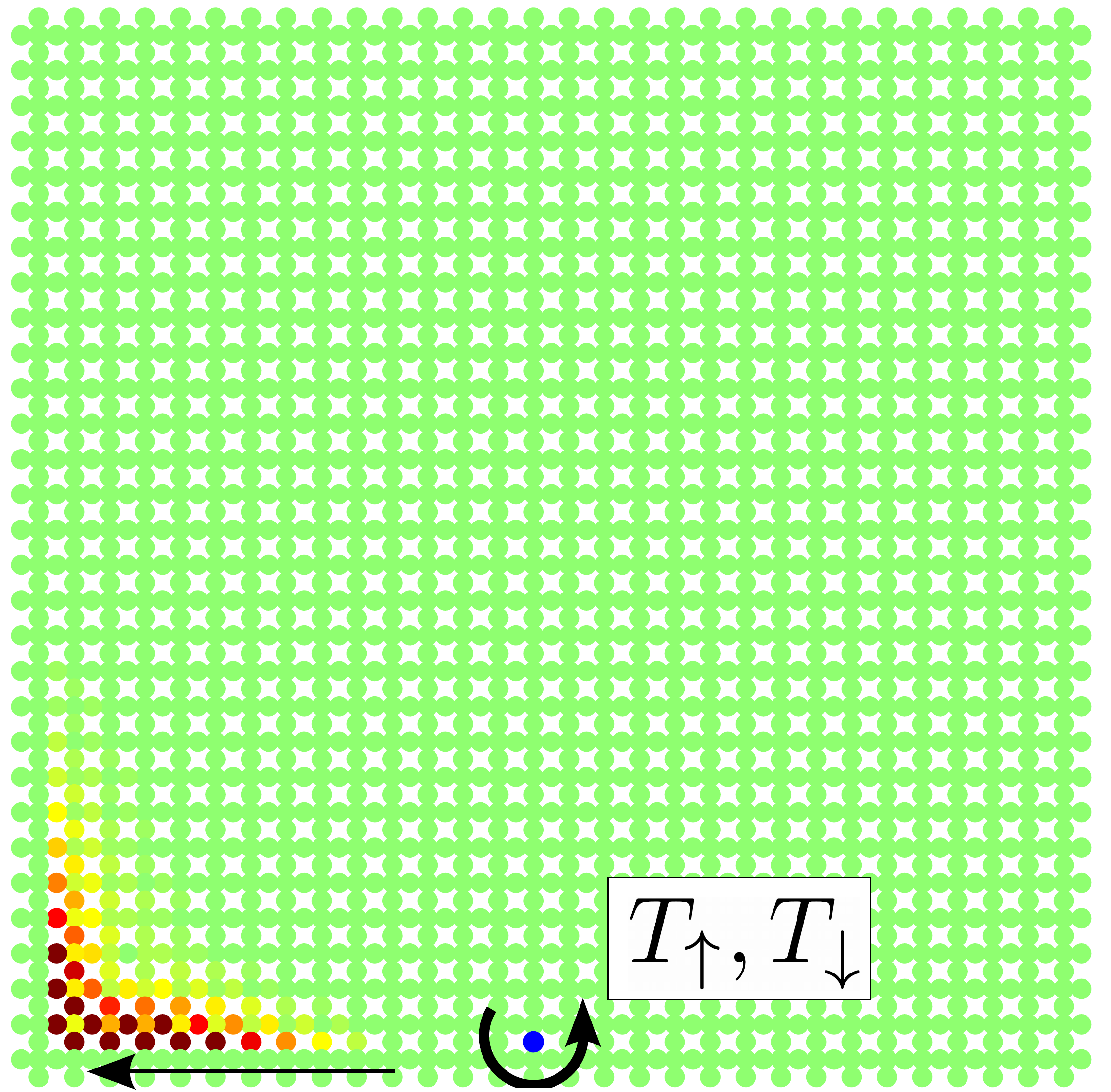}}
\subfloat[$\tau = 600$]{
	\includegraphics[keepaspectratio,width=0.33\columnwidth]{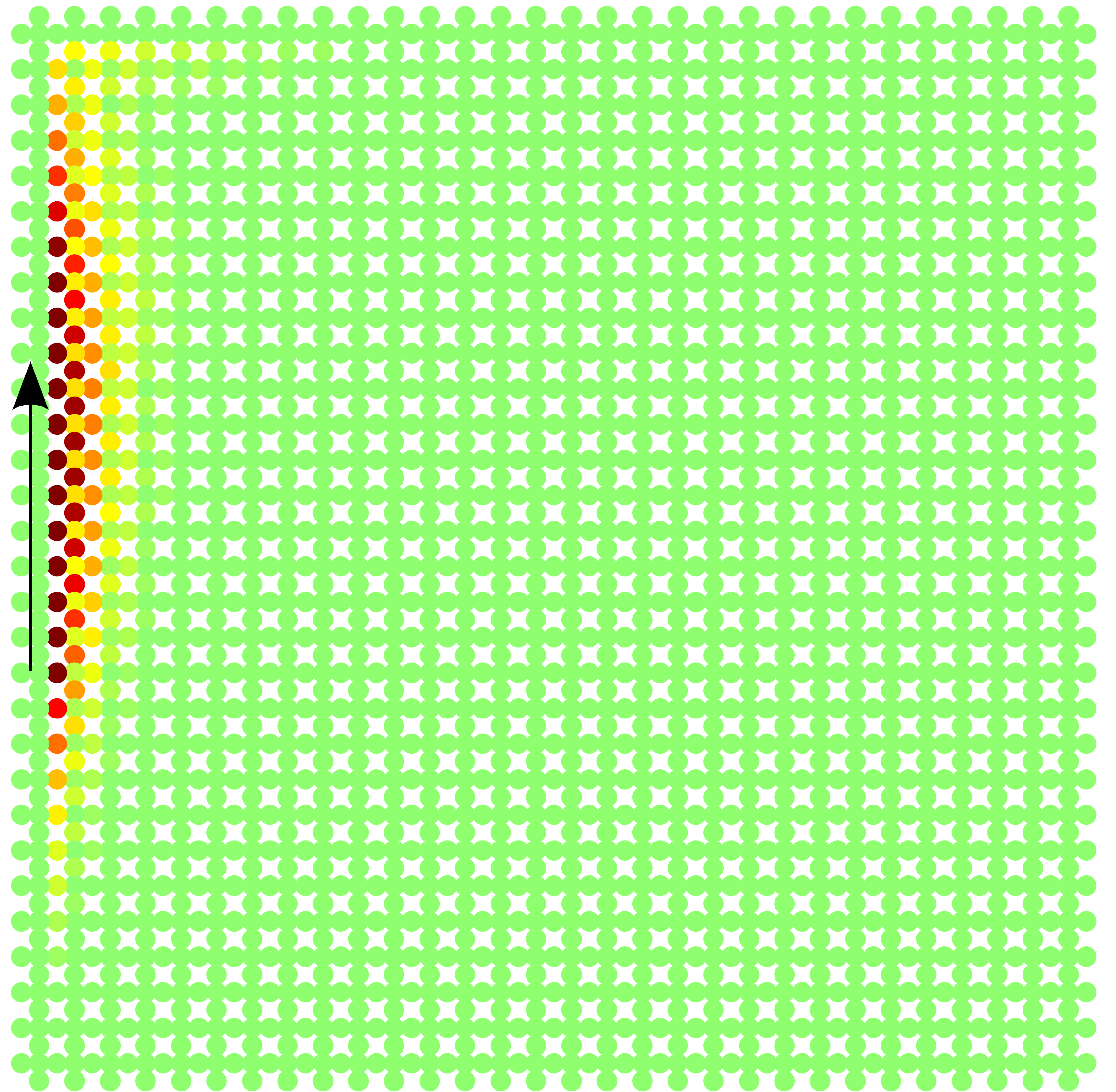}}
\subfloat[$\tau = 900$]{
	\includegraphics[keepaspectratio,width=0.33\columnwidth]{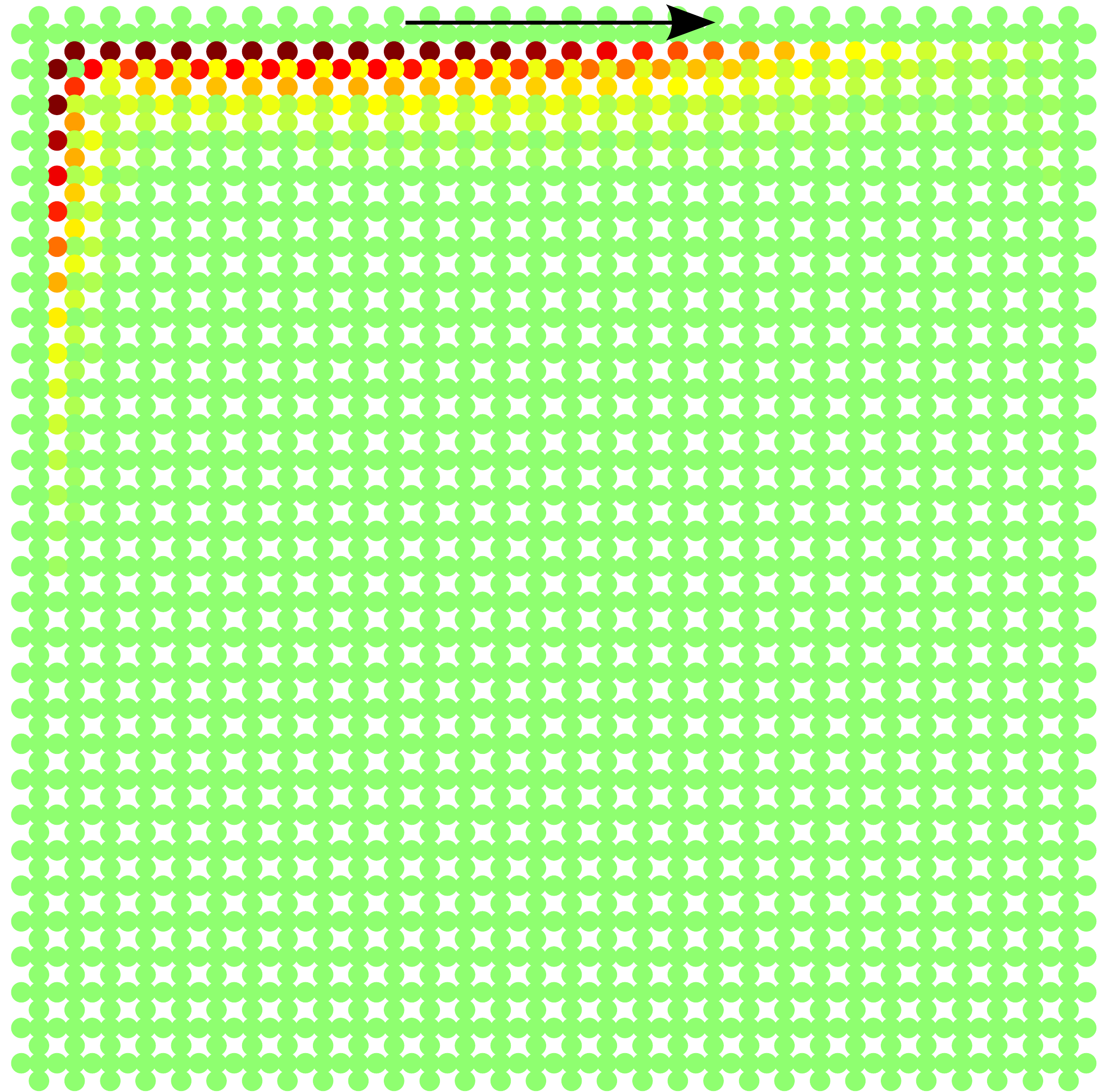}} \\
\subfloat[$\tau = 200$]{
	\includegraphics[keepaspectratio,width=0.33\columnwidth]{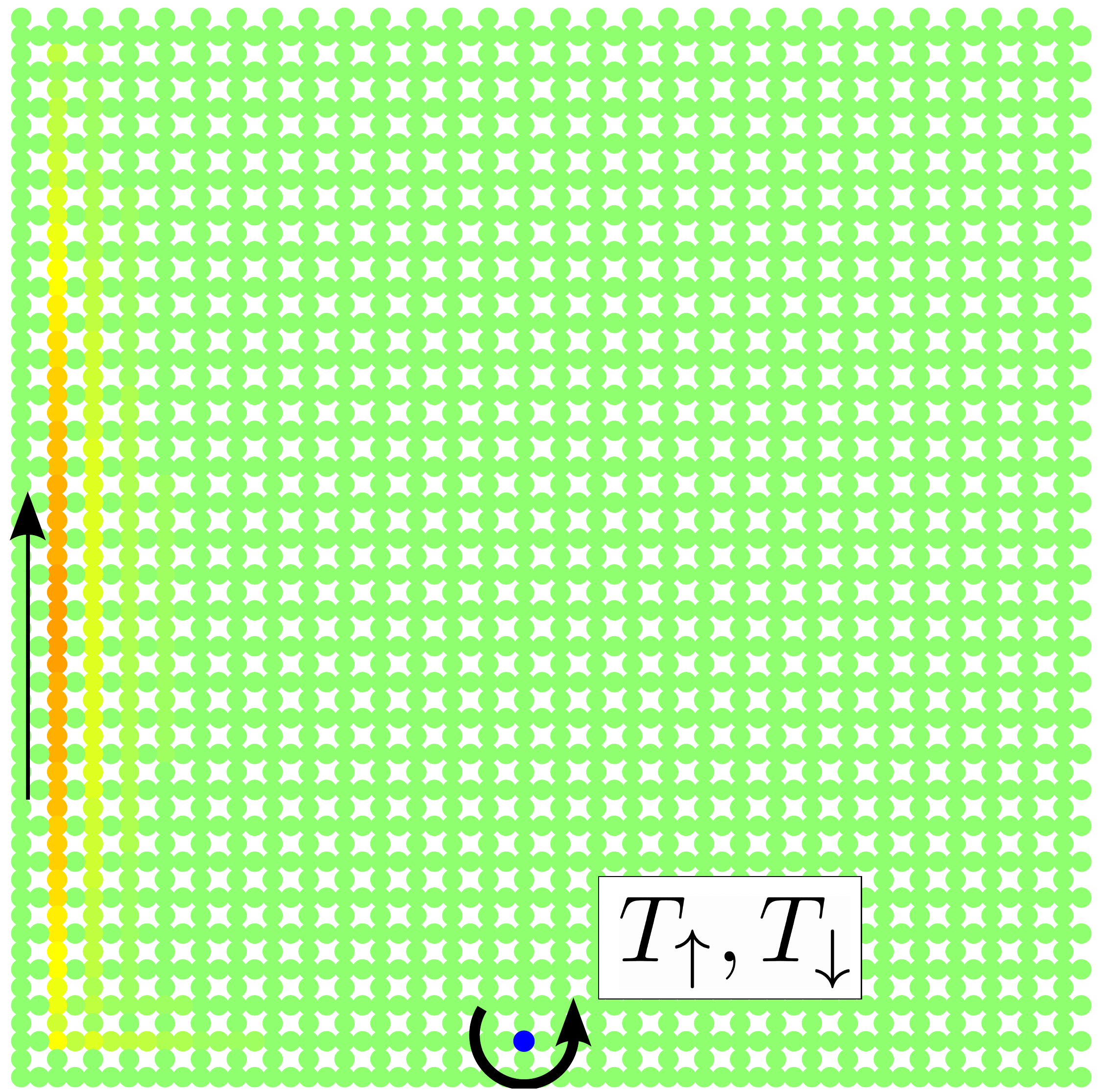}}
\subfloat[$\tau = 400$]{
	\includegraphics[keepaspectratio,width=0.33\columnwidth]{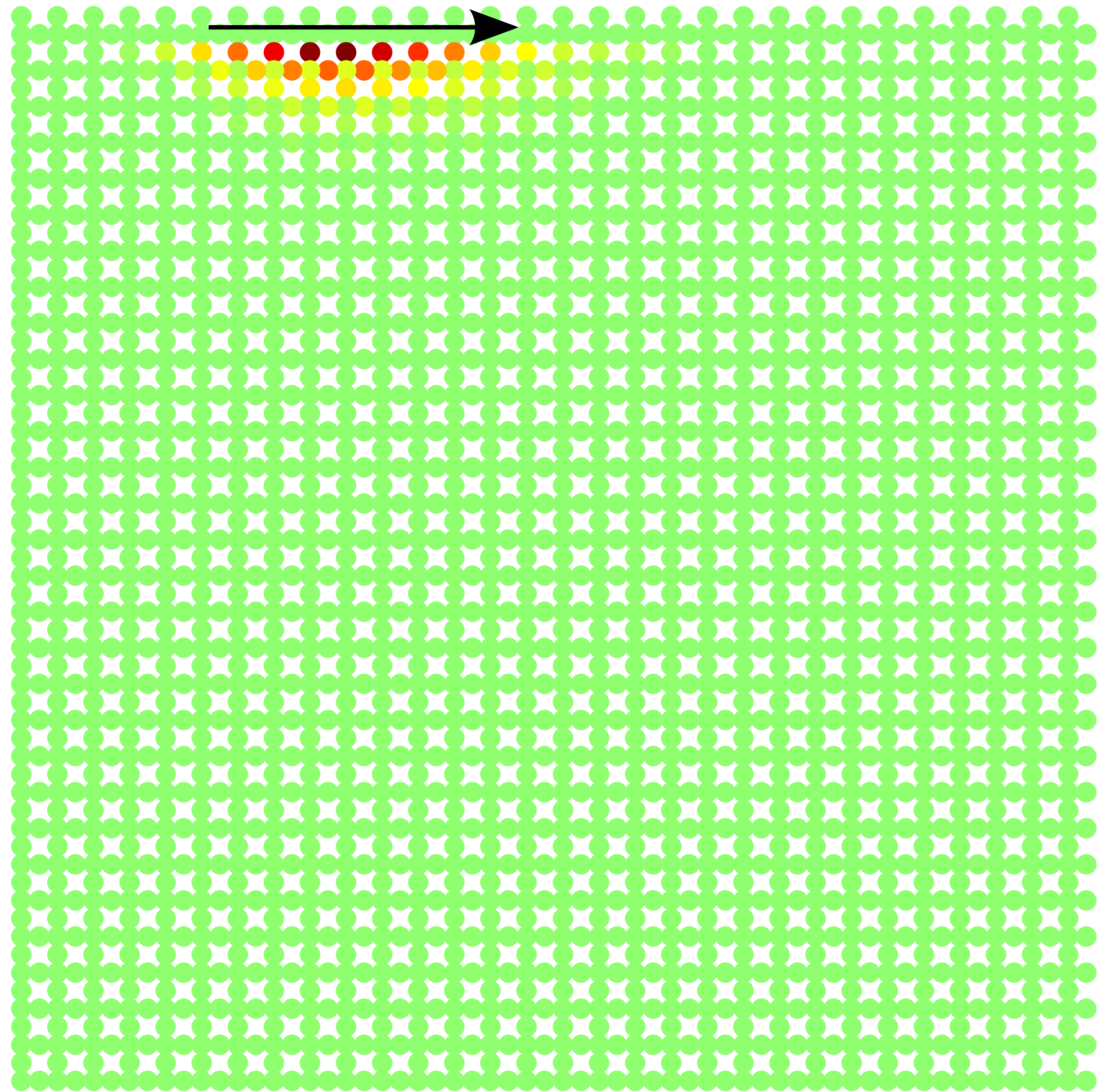}}
\subfloat[$\tau = 600$]{
	\includegraphics[keepaspectratio,width=0.33\columnwidth]{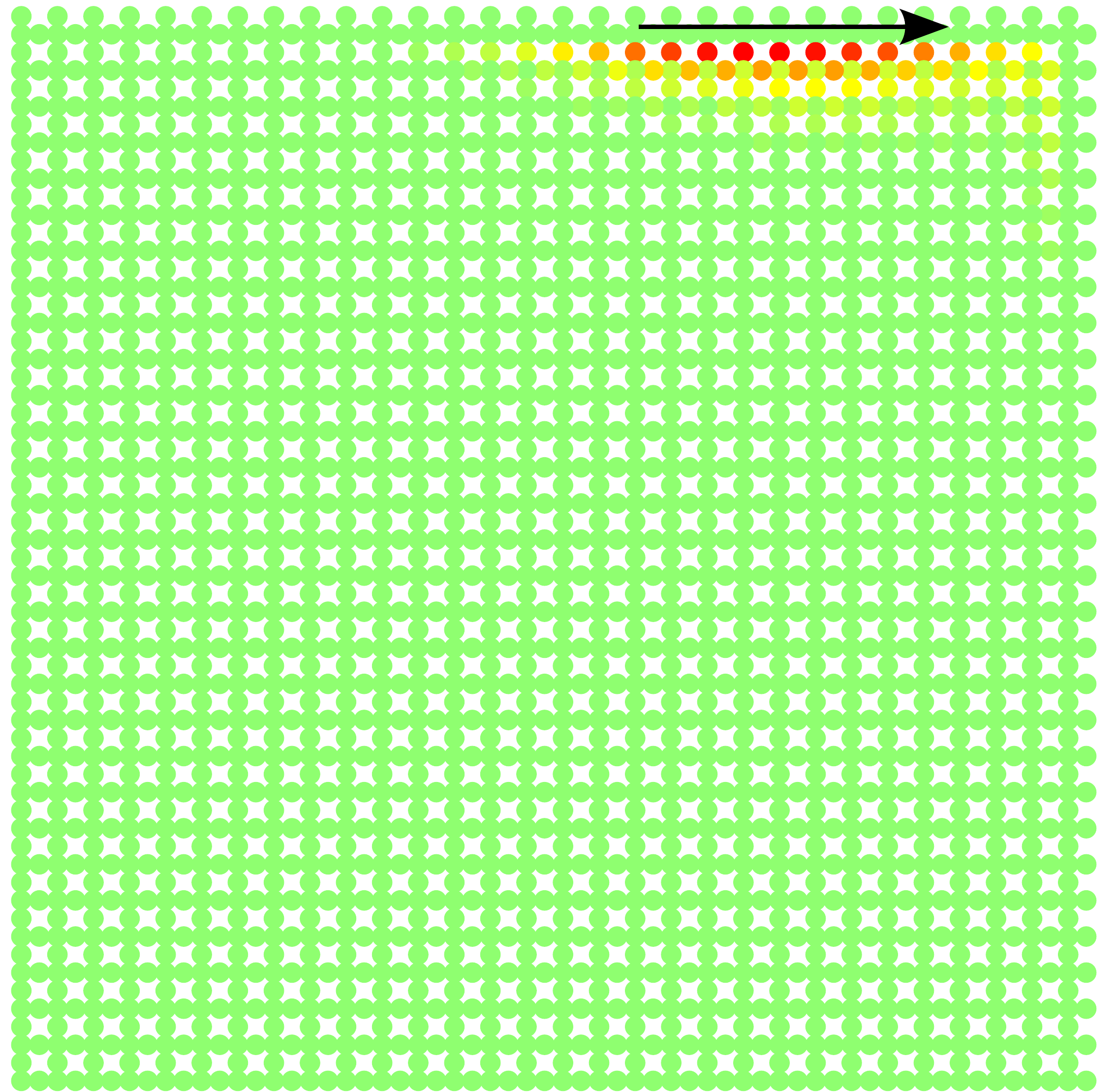}}
	\caption{RMS displacement at each lattice site for different moments in time $\tau$. (a-c) Clockwise propagation with one edge type.  (d-f) Clockwise propagation with two edge types.   Red denotes the greatest displacement and green denotes none.}
	\label{fig_Lieb_NS}
\end{figure}

Our final example studies the Lieb lattice.
The bulk dispersion surface of the mechanical Lieb lattice with $k_{nn} = 1$ and $\lambda_{iso} = 0$ is pictured in \sFigref{fig_Lieb_noISO}, projected on the $x\Omega$-plane with the wavenumber $\kappa_x$ normalized as in the previous subsection.  Low frequency waves cannot propagate, due to the presence of ground springs.  The addition of inter layer coupling with $\lambda_{iso} = 0.2$ opens two bandgaps spanning $\Omega \approx 2.0$ to $2.2$ and $\Omega \approx 2.2$ to $2.4$.  

A strip of the Lieb lattice is analyzed in the same way as the hexagonal lattice, with the disks at the boundary fixed. 
Note that there are two types of edges arising in the Bloch analysis of a strip. The first edge corresponds to the edge of a unit cell closer to the 
$x$-axis in the schematic in Fig.~\ref{fig_Lieb_cell}, while the second type corresponds to the opposite edge of the unit cell. 
Figure~\subref*{fig_Lieb_edge} displays the corresponding dispersion diagram. 
There are $4$ distinct modes localized at the boundary which span the bulk bandgaps.  They correspond to the left and right propagating modes on the two distinct boundaries. 

For numerical simulations, first a lattice is considered with all boundaries identical.
The edge described by the upper end of the strip, i.e. positive $y$ in \sFigref{fig_Lieb_cell}, is used to demonstrate the excitation of a TPBM in a simulation of the Lieb lattice.  The unit cell is tessellated into a 30 by 30 cell lattice and sites are added where needed to make all the boundaries of the same construction.  Then, the outer layer of cells is fixed.  The same forcing input as for the hexagonal lattice is used except $\Omega = 2.1$.  With $T_\uparrow = \cos(\Omega\tau)$ and $T_\downarrow = \sin(\Omega\tau)$ a clockwise-propagating mode is excited, but if $T_\downarrow = -\sin(\Omega\tau)$, a counter-clockwise propagating wave is excited. Figure \ref{fig_Lieb_NS}(a-c) depicts the results, and the waves are seen to transition between edges without reflections.

Next, we present numerical simulations of a wave propagating from  one boundary type to another (\Figref{fig_Lieb_NS}(d-f)).  The same forcing input is used as in the aforementioned simulation, and the simulation domain is a 30 by 30 cell lattice with the outer layer of cells fixed.  The wave packet propagates on both edge types at different velocities, and exhibits no back-scattering when transitioning from one edge to the other.  In \Figref{fig_Lieb_NS}(e-f) the wave packet looks smaller than in \Figref{fig_Lieb_NS}(a-c) because the ``smooth'' edge type is not as dispersive as the other, and the wave has not had as much time to disperse. Indeed, this behavior is consistent with the trends observed in the band diagram, where there are two group velocities at 
each frequency in the bulk bandgap region. Each mode corresponds to a localized wave on a particular boundary type and they are all unidirectional. As the wave traverses from one 
boundary type to the other, there is thus a change in group velocity as the wave transitions from one branch to another.


\begin{figure}
\centering
\subfloat[]{
	\includegraphics[keepaspectratio,width=0.5\columnwidth]{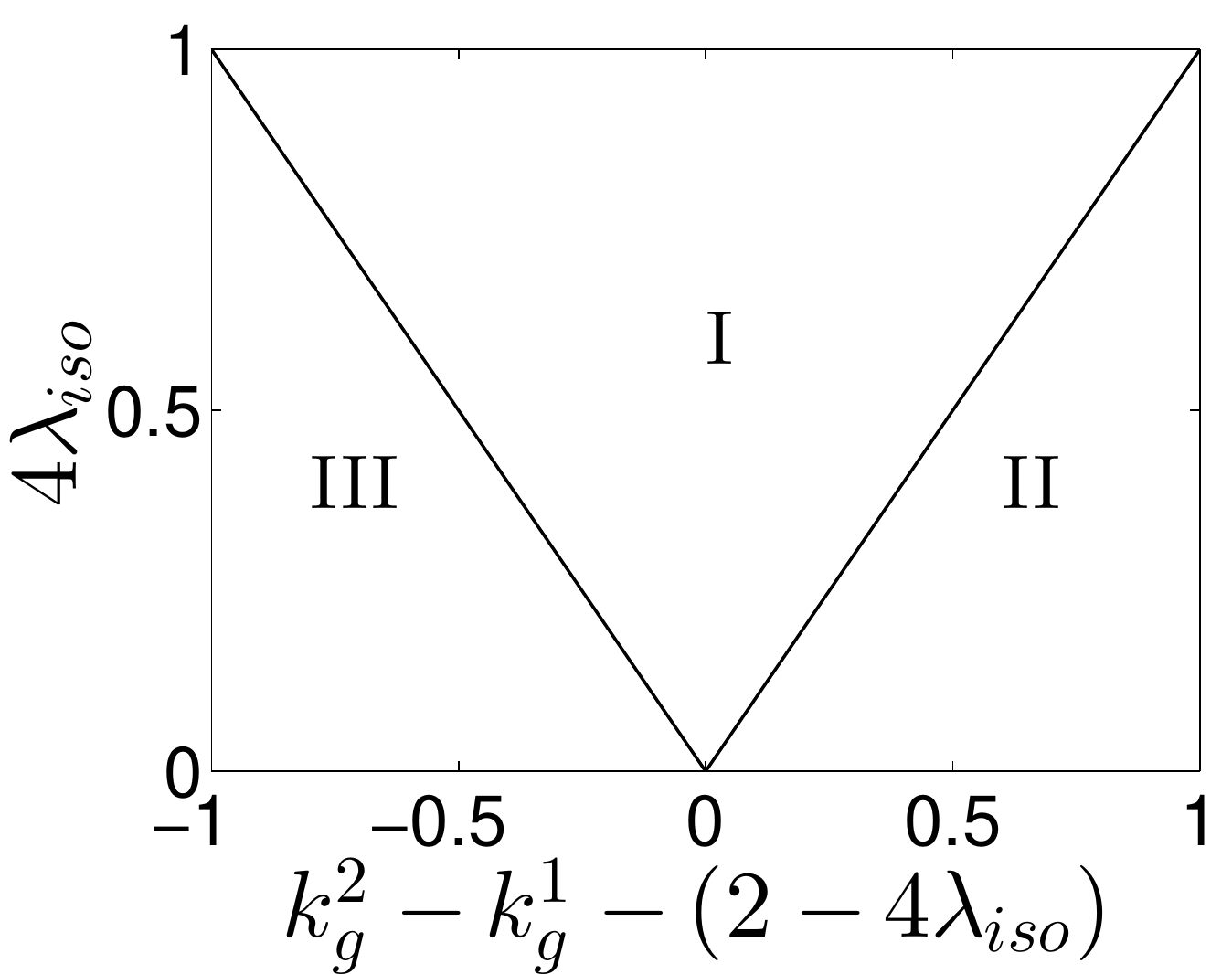}} 
\subfloat[Phase I]{
	\includegraphics[keepaspectratio,width=0.5\columnwidth]{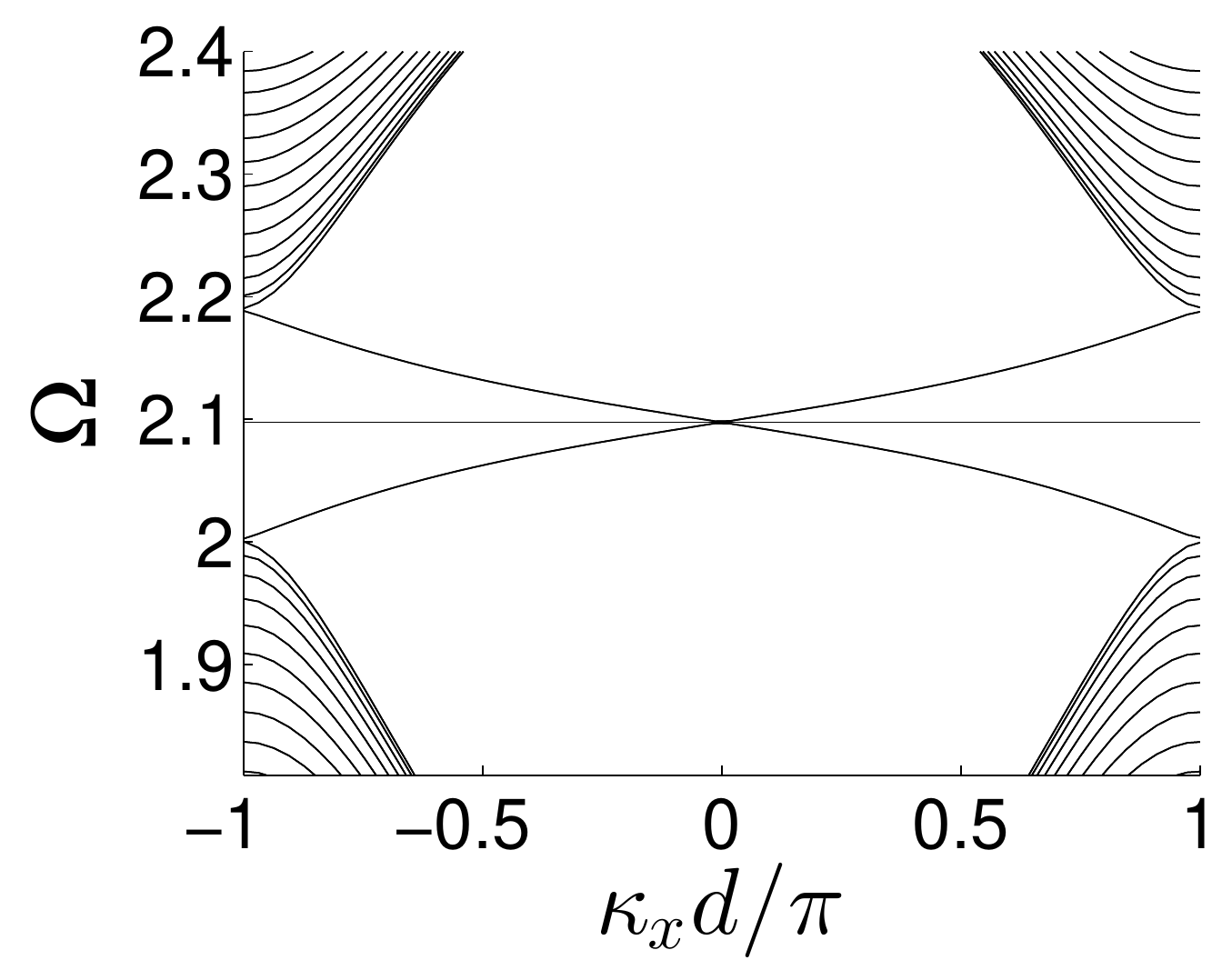}} \\
\subfloat[Phase II]{
	\includegraphics[keepaspectratio,width=0.5\columnwidth]{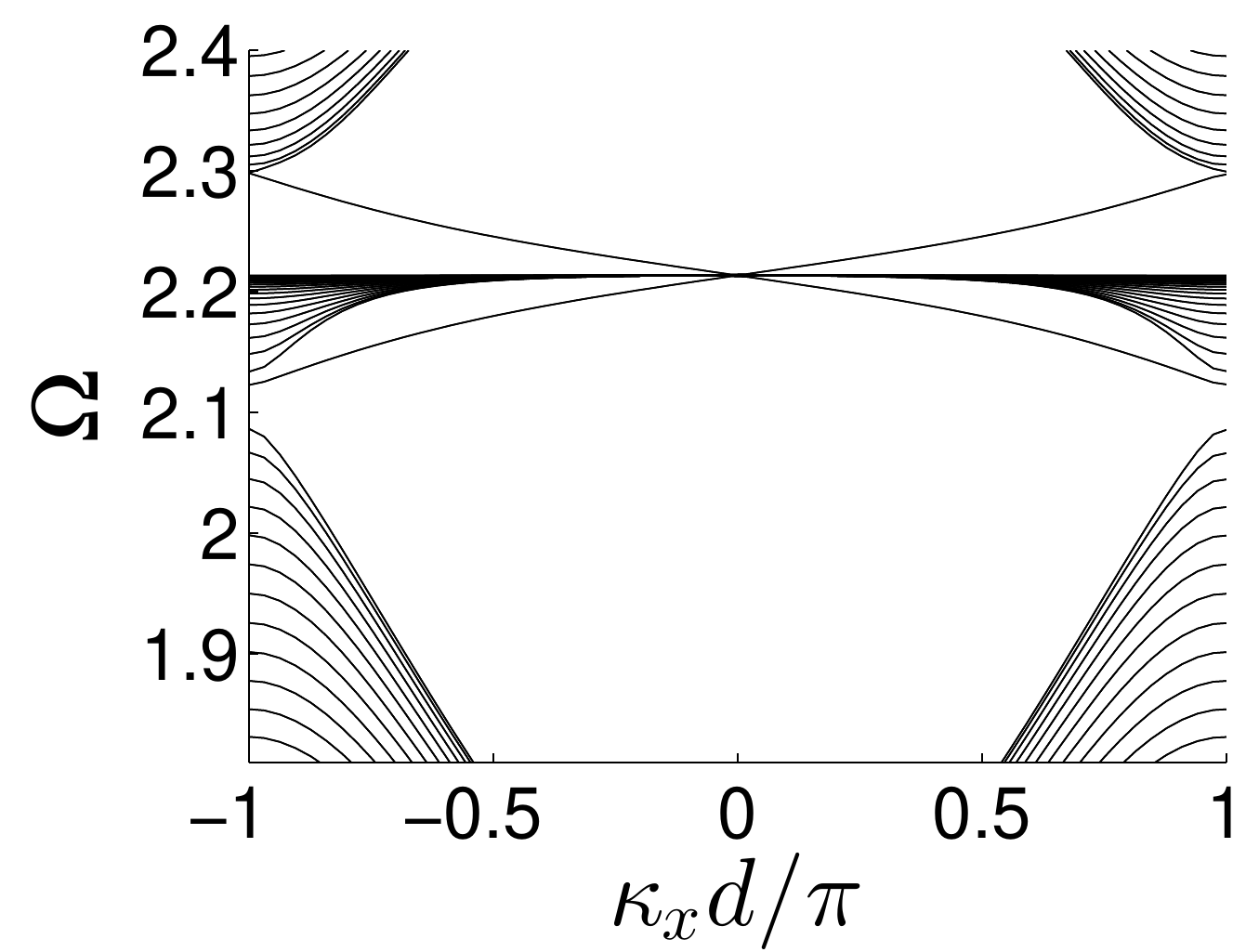}}
\subfloat[Phase III]{
	\includegraphics[keepaspectratio,width=0.5\columnwidth]{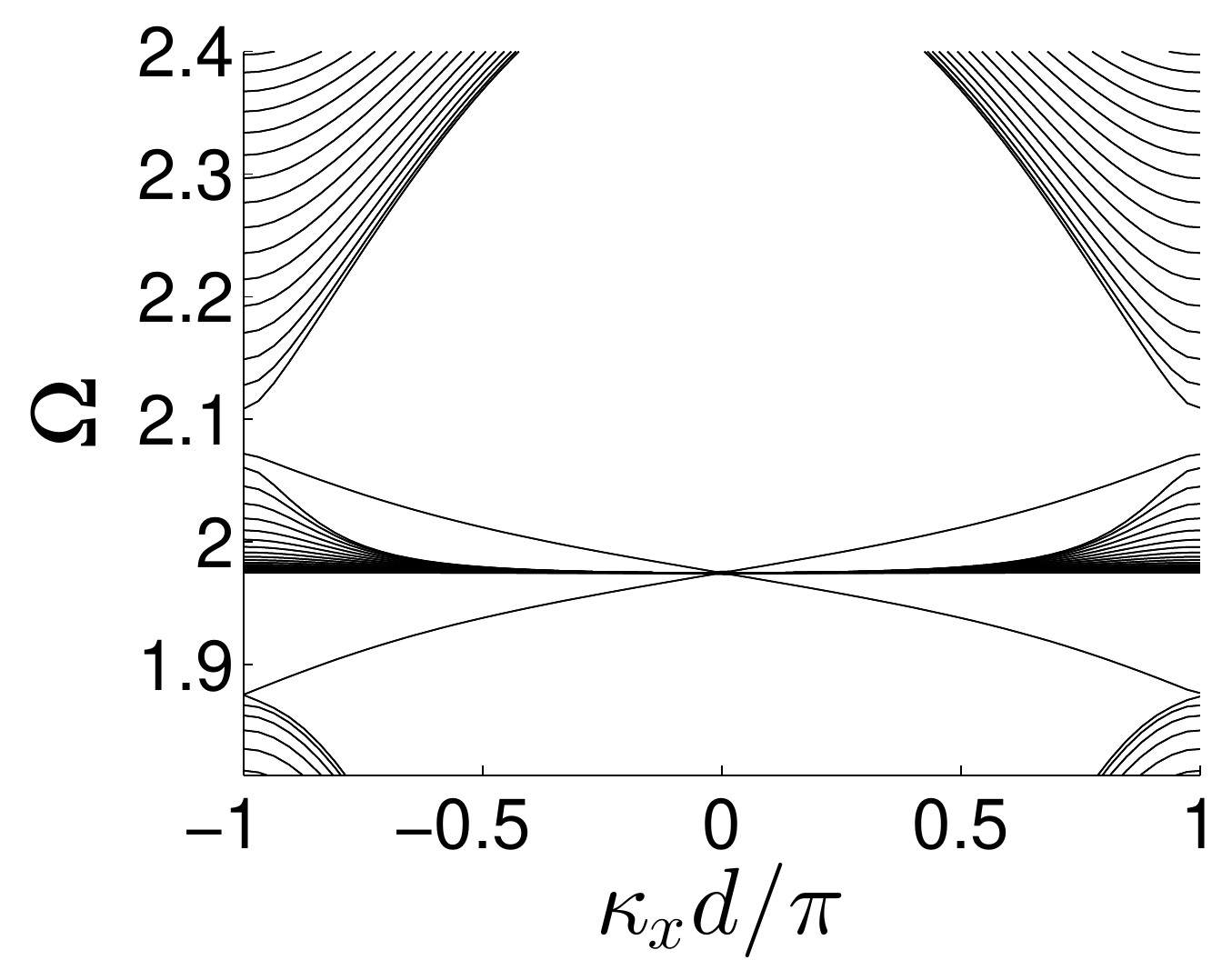}} 
	\caption{(a) Phase diagram with ground spring and interlayer spring stiffness parameters, showing three distinct topological phases. 
	(b-d) Close-up view of band diagrams corresponding to each of the phases. }
	\label{fig_kg_phase}
\end{figure}
\subsection{Topological phase transitions by varying ground spring stiffness}

We finally discuss the effect of ground springs on the band structure. We label the sites having $4$ and $2$ nearest neighbors by indices $1$ and $2$, respectively. 
There are two sets of ground springs, $k_{g1}$ and $k_{g2}$ connecting the site types $1$ and $2$ with the ground, 
respectively. As mentioned earlier, modifying both the ground springs by the same amount has the effect of translating the entire band along the frequency axis without altering
their topology. The values of ground springs which correspond to a flat band, an exact analogue of the quantum mechanical case, satisfy the constraint $k_{g1} - k_{g2} = 2k_{nn} - 4 \lambda_{iso}$. 
We thus vary $k_{g2}$ keeping $k_{g1}$ fixed, 
and observe the effect on the band structure by computing the spin Chern numbers~\cite{fukui2005chern,beugeling2012topological}. 
\newtext{The Chern numbers $C_+$ and $C_-$ of the top and bottom bands related as $C_+ = - C_-$ and the spin Chern number is given by $C_s  = C_+ - C_-$. }
Figure~\ref{fig_kg_phase} illustrates the phase diagram 
of the system as a function of the relative ground stiffness and the interlayer coupling stiffness. These stiffness values are normalized by the in-plane stiffness $k_{nn}$. 

There are three regions (I,II,III) in the phase diagram, with phase boundaries between them. The central region (I) has TPBMs between the lower and the upper bands, and is topologically 
equivalent to the case studied earlier having a flat band at the center. \newtext{The bands have spin Chern number $C_s = (-2,0,2)$.} The region on the left (III) has TPBMs only between the middle band and the lower band, while the region on the right (II)
has TPBMs between the top and middle bands. \newtext{These regions have spin Chern numbers $C_s = (-2,2,0)$ and $(0,-2,2)$, respectively. Indeed the change in band topology is reflected in the change in
their respective Chern numbers. }
Thus the presence of edge modes is insensitive to small changes in the ground springs, as long as they remain within the phase boundaries, thereby illustrating the robustness to 
imperfections in material properties. It also shows the parameter space over which these TPBMs can be achieved and provides guidelines for designing lattices. 
Figure~\ref{fig_kg_phase} illustrates the band diagrams at each of the phases. In these calculations the interlayer spring stiffness is fixed at $\lambda_{iso}=0.1$, while the relative strengths
of the ground springs is varied so that $k_{g2}-k_{g1}-(2-4\lambda_{iso})$ takes values $\{0,0.5,-0.5\}$ in Figs.~\ref{fig_kg_phase}(b)-(d). 
Note that as the ground stiffness deviates from the value $k_{g2}-k_{g1} = 2 - 4\lambda_{iso}$, the center band is not flat anymore. The center band grows and touches either the top or 
the lower band depending on the value of $k_{g2}-k_{g1}$, at which point a phase transition occurs.

To observe this phase transition from the structure of the
dynamical matrix, we apply the inverse transform $(\bH = \bU \bD_{H} \bU^{\dag})$ to a lattice with arbitrary ground springs to recover the equivalent quantum Hamiltonian. This transform
simplifies the eigenvalue analysis as the top and bottom layers are effectively decoupled. 
Applying Bloch reduction on a unit cell results in 
the characteristic equation $(\alpha - k_{g2} +k_{g1} + 2 - 4\lambda_{iso})(\alpha^2 - 16\lambda_{iso}^2 ) = 0$ for the eigenvalues $\alpha$ at the point $\kappa_x = \kappa_y = \pi/2$. A topological phase transition occurs 
only when the bands touch, ie., when there are degenerate eigenvalues. Indeed, these degenerate values arise when $\alpha = k_{g2} - k_{g1}- 2+4\lambda_{iso} = \pm 4 \lambda_{iso}$, which are the phase 
transition boundaries in Fig.~\ref{fig_kg_phase}.  
We finally remark on the tunability of our lattices by varying the ground springs. These phase transitions give a way to modify the behavior of the lattice at a particular frequency from
unidirectional propagation along the edge to bulk modes to no propagation along the edges and the interior. 

\section{Conclusions}\label{ConcSection}

A framework for generating a family of mechanical systems that exhibit TPBMs is derived and demonstrated numerically. 
A mechanical analogue of a quantum mechanical system that exhibits edge modes is obtained by an appropriate transformation and this analogue defines the lattice network topology. 
The phase boundaries of the topological phase in the vicinity of the parameters defining the mechanical analogue are demonstrated to show the robustness to mechanical imperfections. 
The hexagonal and Lieb lattice examples show the efficacy of the methodology and numerical simulations reinforce it.  Excitation polarization is shown to determine the direction of energy propagation.  The Lieb lattice is used to highlight that the type of edge influences the boundary wave propagation. Finally, topological phase transitions are obtained by modifying
the ground springs, thereby illustrating the potential for tunable wave propagation in these lattices. Future work entails extension of these concepts to incorporate general mechanical elements.


\bibliographystyle{unsrt}
\bibliography{paper}

\section*{Appendix 1: Equations of Motion}
\label{Appendix1}
We present the complete equations of motion for the disks in a unit cell for both the lattices considered in this work. Consider an infinite lattice divided into unit cells, and each
unit cell is indexed by a pair of integers $(i,j)$, denoting its location with respect to a fixed coordinate system. 
Let $I$ be the rotational inertia of each disk and let $\theta_{i,j}^{a,b}$ denote the angular displacement of a disk. 
The two subscripts $(i,j)$ identify the unit cell and and the two super-scripts identify the location of the disk within the unit cell $(i,j)$. 
The first super-script index denotes its location in the $xy$-plane while the second 
super-script index denotes whether the disk is in top (t) or bottom (b) layer. 

\subsection*{Hexagonal lattice}
\begin{figure}
\centering
\includegraphics[scale=0.2]{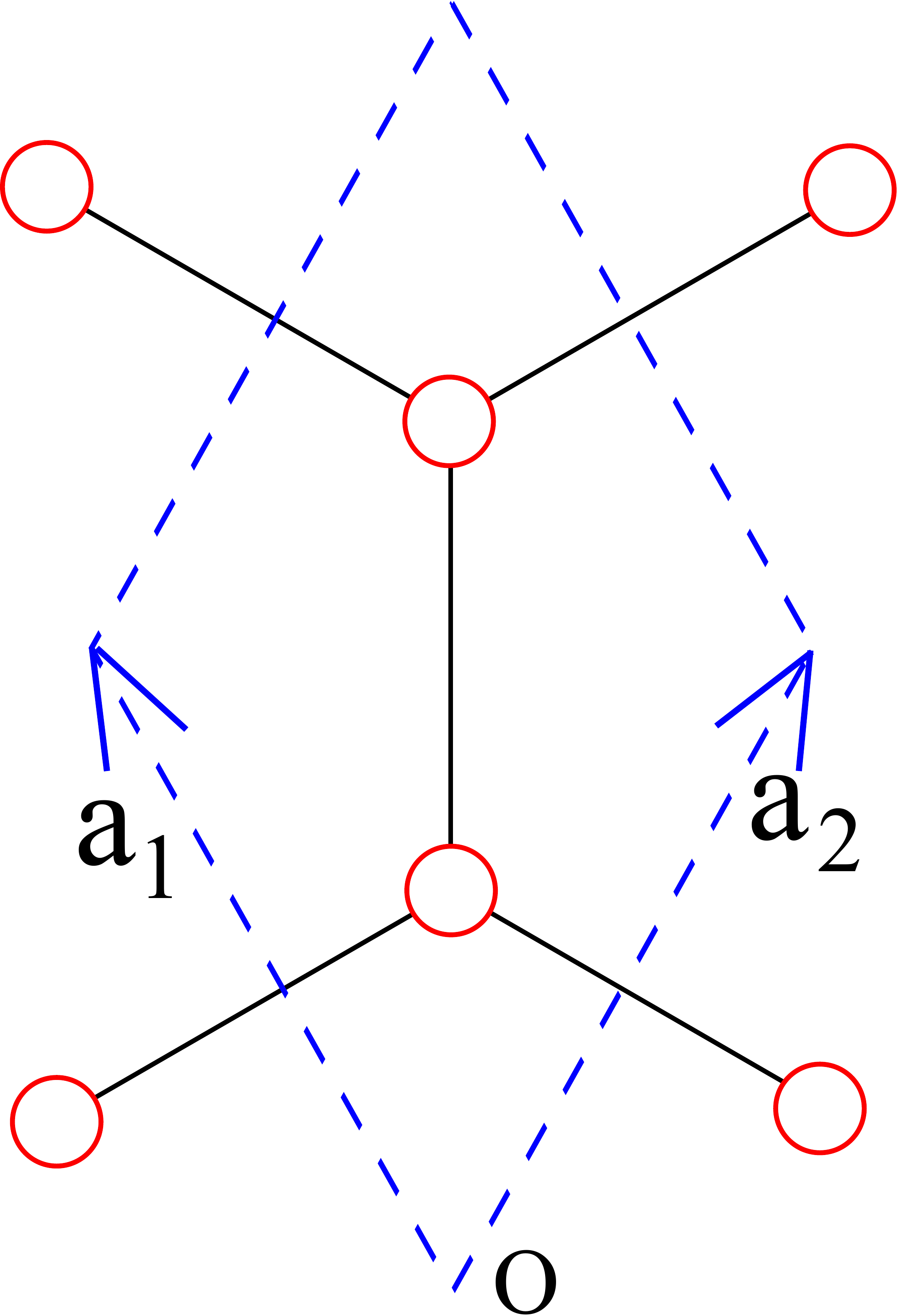}
\caption{Schematic of a hexagonal unit cell. The top layer is shown, along with index of the disks. }
\label{HexIndexUnit}
\end{figure}
Figure~\ref{HexIndexUnit} displays the top layer of a unit cell of the hexagonal lattice. Each layer has two disks, indexed as illustrated. The equations of motion of the four disks 
of the unit cell are
\begin{subequations}
\begin{multline}
I \ddot{\theta}_{i,j}^{1,t}  = k_{nn} \left(\theta_{i,j}^{2,t} + \theta_{i,j-1}^{2,t} + \theta_{i-1,j}^{2,t} - 3\theta_{i,j}^{1,t} \right) - 6 k_{iso}\theta_{i,j}^{1,t} \\
								+ k_{iso}\left( \theta_{i+1,j}^{1,b} + \theta_{i-1,j+1}^{1,b} + \theta_{i,j-1}^{1,b} \right) \\
								- k_{iso}\left( \theta_{i,j+1}^{1,b} + \theta_{i-1,j}^{1,b} + \theta_{i+1,j-1}^{1,b} \right) 
\end{multline}
\begin{multline}
I \ddot{\theta}_{i,j}^{2,t}  = k_{nn} \left(\theta_{i,j}^{1,t} + \theta_{i,j+1}^{1,t} + \theta_{i+1,j}^{1,t} - 3\theta_{i,j}^{2,t} \right) - 6 k_{iso}\theta_{i,j}^{2,t} \\
								+ k_{iso}\left( \theta_{i,j+1}^{2,b} + \theta_{i-1,j}^{2,b} + \theta_{i+1,j-1}^{2,b} \right) \\
								- k_{iso}\left( \theta_{i+1,j}^{2,b} + \theta_{i-1,j+1}^{2,b} + \theta_{i,j-1}^{2,b} \right) 
\end{multline}
\begin{multline}
I \ddot{\theta}_{i,j}^{1,b}  = k_{nn} \left(\theta_{i,j}^{2,b} + \theta_{i,j-1}^{2,b} + \theta_{i-1,j}^{2,b} - 3\theta_{i,j}^{1,b} \right) - 6 k_{iso}\theta_{i,j}^{1,b} \\
								+ k_{iso}\left( \theta_{i+1,j}^{1,t} + \theta_{i-1,j+1}^{1,t} + \theta_{i,j-1}^{1,t} \right) \\
								- k_{iso}\left( \theta_{i,j+1}^{1,t} + \theta_{i-1,j}^{1,t} + \theta_{i+1,j-1}^{1,t} \right) 
\end{multline}
\begin{multline}
I \ddot{\theta}_{i,j}^{2,b}  = k_{nn} \left(\theta_{i,j}^{1,b} + \theta_{i,j+1}^{1,b} + \theta_{i+1,j}^{1,b} - 3\theta_{i,j}^{2,b} \right) - 6 k_{iso}\theta_{i,j}^{2,b} \\
								+ k_{iso}\left( \theta_{i,j+1}^{2,t} + \theta_{i-1,j}^{2,t} + \theta_{i+1,j-1}^{2,t} \right) \\
								- k_{iso}\left( \theta_{i+1,j}^{2,t} + \theta_{i-1,j+1}^{2,t} + \theta_{i,j-1}^{2,t} \right) 
\end{multline}
\end{subequations}

\subsection*{Lieb lattice}

\begin{figure}
\centering
\includegraphics[scale=0.2]{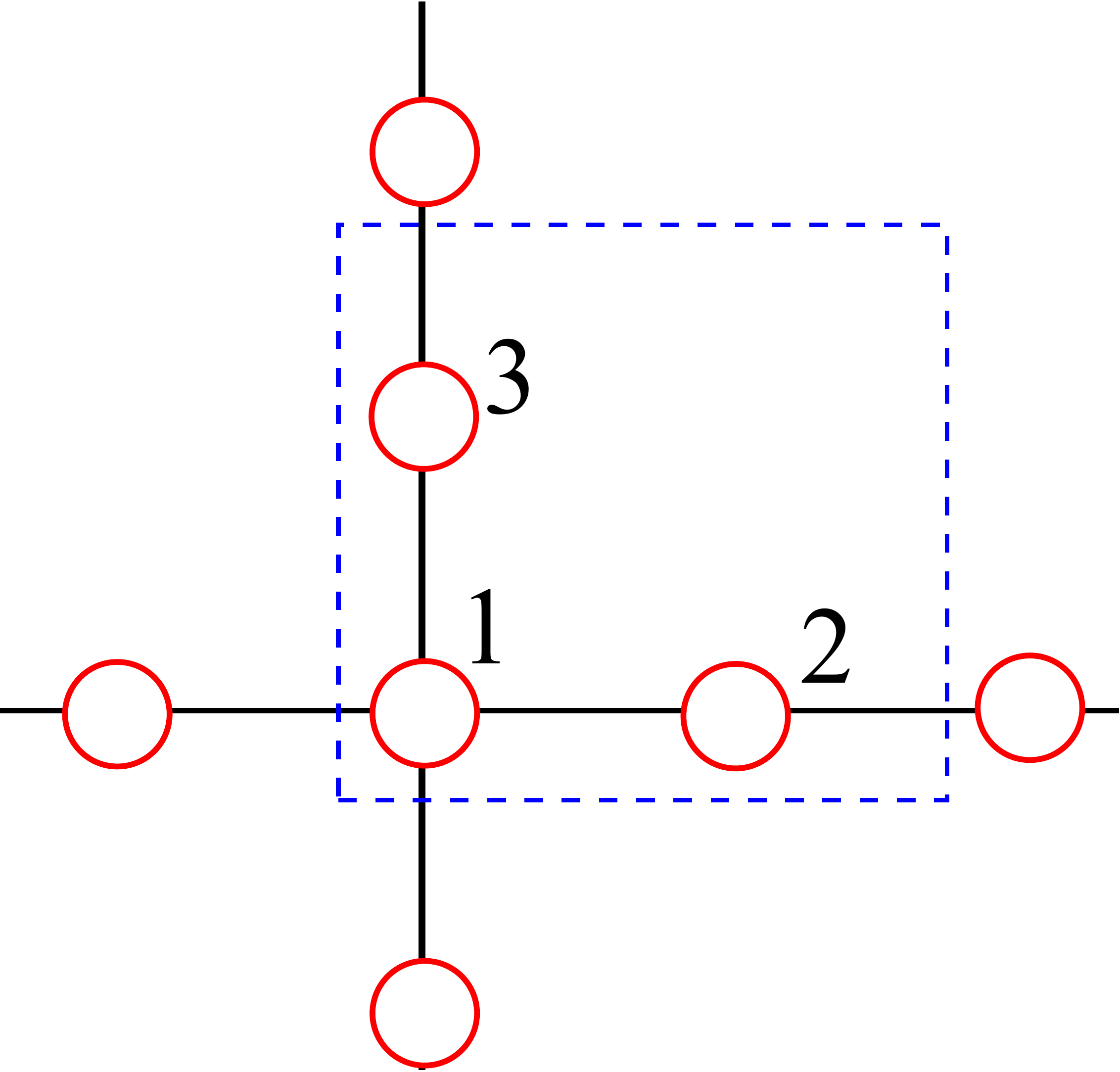}
\caption{Schematic of a Lieb lattice unit cell. The top layer is shown, along with index of the disks. }
\label{LiebIndexUnit}
\end{figure}

Figure~\ref{LiebIndexUnit} displays the top layer of a unit cell of the Lieb lattice. Each layer has three disks, indexed as illustrated. 
Let $k_g = \max(2k_{nn} + 4\lambda_{iso},4k_{nn})$ be the additional term required to shift all the bands upward. 
The equations of motion of the six disks of the unit cell are
\begin{subequations}
\begin{multline}
I \ddot{\theta}_{i,j}^{1,t}  =  k_{nn} \left( \theta_{i,j}^{2,t} + \theta_{i,j}^{3,t} + \theta_{i,j-1}^{3,t} + \theta_{i-1,j}^{2,t}  \right) 
																		- k_{g} \theta_{i,j}^{1,t},
\end{multline}
\begin{multline}
I \ddot{\theta}_{i,j}^{2,t}  =  k_{nn} \left( \theta_{i,j}^{1,t} + \theta_{i+1,j}^{1,t}  \right) 
																		- k_{g} \theta_{i,j}^{2,t} \\ 
																		+ k_{iso} \left( \theta_{i+1,j}^{3,b} - \theta_{i,j}^{3,b} + \theta_{i,j-1}^{3,b} - \theta_{i+1,j-1}^{3,b} \right) 
\end{multline}
\begin{multline}
I \ddot{\theta}_{i,j}^{3,t}  =  k_{nn} \left( \theta_{i,j}^{1,t} + \theta_{i,j+1}^{1,t}  \right) 
																		- k_{g} \theta_{i,j}^{3,t} \\ 
																		+ k_{iso} \left( \theta_{i-1,j+1}^{2,b} - \theta_{i-1,j}^{2,b} + \theta_{i,j}^{2,b} - \theta_{i,j+1}^{2,b} \right) 
\end{multline}
\begin{multline}
I \ddot{\theta}_{i,j}^{1,b}  =  k_{nn} \left( \theta_{i,j}^{2,b} + \theta_{i,j}^{3,b} + \theta_{i,j-1}^{3,b} + \theta_{i-1,j}^{2,b}  \right)  
																		- k_{g} \theta_{i,j}^{1,b},
\end{multline}
\begin{multline}
I \ddot{\theta}_{i,j}^{2,b}  =  k_{nn} \left( \theta_{i,j}^{1,b} + \theta_{i+1,j}^{1,b}  \right) 
																		- k_{g} \theta_{i,j}^{2,b} \\ 
																		+ k_{iso} \left( \theta_{i+1,j}^{3,t} - \theta_{i,j}^{3,t} + \theta_{i,j-1}^{3,t} - \theta_{i+1,j-1}^{3,t} \right) 
\end{multline}
\begin{multline}
I \ddot{\theta}_{i,j}^{3,b}  =  k_{nn} \left( \theta_{i,j}^{1,b} + \theta_{i,j+1}^{1,b}  \right) 
																		- k_{g} \theta_{i,j}^{3,b} \\ 
																		+ k_{iso} \left( \theta_{i-1,j+1}^{2,t} - \theta_{i-1,j}^{2,t} + \theta_{i,j}^{2,t} - \theta_{i,j+1}^{2,t} \right) 
\end{multline}
\end{subequations}
Note that the disks indexed $1$ and $\{2,3\}$ have different number of neighbors and hence ground springs are required to ensure that 
all the bands are shifted by the same amount. The ground stiffness is required on either the $1$ disks or on the $\{2,3\}$ depending
on the relative magnitudes of $k_{nn}$ and $k_{iso}$. Its value is $4k_{iso}-2k_{nn}$ on disk $1$ if $2k_{iso} > k_{nn}$ or $2k_{nn}-4k_{iso}$
on disks $\{2,3\}$ in each unit cell.

\section*{Appendix 2: Bloch Wave Analysis of a Strip}
\label{Appendix2}

A convenient way to check for the existence of TPBMs is to calculate the band diagram for the 1D, periodic system created in the following way.  The unit cell is tessellated a finite number of times along one lattice vector to make a strip, and then made periodic along the other lattice vector.  The result is a system with two parallel boundaries. TPBMs, if they exist in a specific structure, will manifest in the frequency range of the bulk bandgap due to the bulk-edge correspondence principle~\cite{jackiw1976solitons,hasan2010colloquium}.  Furthermore, the deformation associated with these modes, i.e. the eigenvectors, will be localized at one of the boundaries.  Indeed, there is some redundancy in solving for edge modes this way, as an infinite half space would describe only one edge.  However, this method requires only a small modification from a 2D Bloch analysis implementation and is therefore quite convenient and practical.  

For the results shown in this article a strip of the hexagonal lattice is made by repeating 27 unit cells along $\bm d_1$ and then made periodic along $\bm d_2$ by assuming the traveling wave solution.  The disks of the unit cells at each end of the strip are rigidly fixed, so there are 25 unit cells elastically connected to two rigid ``walls''.  Similarly, to analyze another type of edge, the ends of the strip can be left free.  For the Lieb lattice the strip is also made by repeating 27 unit cells (\sFigref{fig_Lieb_cell}) along $\bm d_1$, rigidly fixing the unit cells at each end of the strip, and making the strip periodic along $\bm d_2$.

\end{document}